\documentclass{article}
\usepackage{appendix}
\usepackage[a4paper, total={6in, 8in}]{geometry}

\usepackage{natbib}
\bibpunct[, ]{(}{)}{;}{a}{,}{,}

\providecommand{\keywords}[1]
{
  \small	\noindent
  \textbf{Keywords:} #1
}

\usepackage{graphicx} 
\usepackage{bm}
\usepackage{amsmath}
\DeclareMathOperator*{\argmin}{arg\,min}
\DeclareMathOperator*{\argmax}{arg\,max}
\usepackage{amssymb}
\usepackage{xcolor}
\usepackage[font=small, width=0.9\textwidth]{caption}
\usepackage{authblk}
\usepackage{orcidlink}
\usepackage{url}
\usepackage{hyperref}

\usepackage{subfigure}

\usepackage[boxed,linesnumbered]{algorithm2e}

\title{Generating Synthetic Functional Data for Privacy-Preserving GPS Trajectories}

\author[1]{Arianna Burzacchi}
\author[2]{Lise Bellanger}
\author[3]{Klervi Le Gall}
\author[3]{Aymeric Stamm}
\author[1]{Simone Vantini}

\affil[1]{MOX Laboratory, Department of Mathematics, Politecnico di Milano, 20133, Milano, Italy}
\affil[2]{Université Bretagne Sud, Lab-STICC, UMR 6285, Vannes, France}
\affil[3]{Laboratoire de Mathématiques Jean Leray, UMR CNRS 6629, Nantes University, 44322 Nantes, France}

\date{}

\begin{document}

\maketitle
\begin{abstract}
    This research presents FDASynthesis, a novel algorithm designed to generate synthetic GPS trajectory data while preserving privacy. After pre-processing the input GPS data, human mobility traces are modeled as multidimensional curves using Functional Data Analysis (FDA). Then, the synthesis process identifies the $K$-nearest trajectories and averages their Square-Root Velocity Functions (SRVFs) to generate synthetic data. This results in synthetic trajectories that maintain the utility of the original data while ensuring privacy. Although applied for human mobility research, FDASynthesis is highly adaptable to different types of functional data, offering a scalable solution in various application domains.
\end{abstract}


\section{Introduction}\label{sec:intro}

In \citeyear{Sweeney2000Simple}, \citeauthor{Sweeney2000Simple}'s groundbreaking research revealed a critical vulnerability in data privacy: the combination of zip code, date of birth, and gender is sufficient to uniquely identify 87\% of the US population. This discovery underscores how individuals can be easily re-identified based on seemingly harmless data points \citep{Sweeney2000Simple}. Building on \citeauthor{Sweeney2000Simple}'s work, subsequent studies focused on human mobility patterns have further investigated this issue. They demonstrated that, even when explicit personal identifiers are removed, individuals can still be re-identified based on the unique characteristics of their spatiotemporal movement patterns \citep{DeMontjoye2013Unique}. For example, an individual's daily commute, frequented locations, and travel routines create a distinctive trajectory that can be traced back to them. This insight has profound implications for data privacy, revealing that traditional anonymization techniques, which often rely on removing or generalizing personal data, may not be sufficient to protect individual identities. 

With the growing prevalence of personal data disclosure, identity theft, and unauthorized access to personal information, individuals and organizations are increasingly wary of sharing sensitive data. 
The transformation in data management and perception has enhanced the need for laws to control data usage and to preserve the privacy of individuals behind those data. Governments have invested significant effort in proposing and enacting comprehensive privacy legislation, such as the General Data Protection Regulation (GDPR) introduced by the European Union in 2018 as “the toughest privacy and security law in the world” \citep{GDPR}. 
Moreover, traditional methods of data publishing, which involve releasing possibly identifiable data, are increasingly drawing attention due to the privacy risks that they naturally present. There has been a growing recognition of the need for more sophisticated and robust approaches to data anonymization and publishing.
Alongside the growing attention to privacy, there remains a strong interest of industries, academia, and research institutes in collecting or acquiring new representative datasets for data-driven policymaking. Undoubtedly, the utility of large amounts of continuously collected data is undeniable, as it allows us to describe phenomena, highlight peculiarities that might otherwise be challenging to uncover, and especially guide decision-makers toward data-driven and informed solutions.

In this context, synthetic data comes as the solution that balances the privacy-compliance and the usefulness of data. \textit{Synthetic data} is defined as artificial data that is produced from original data by a generative model trained to reproduce the characteristics and structure of the original data. Each observation is generated to mimic the overall behavior of a dataset, thus ensuring data usefulness, but without retaining individual information, thus preserving privacy. 
In the last 10 years, the search for new methods of Synthetic Data Generation (SDG) has grown significantly, and solutions have been proposed for various types of data, from standard tabular data to more complex data structures.
Concerning human mobility, generative methods have been proposed to synthesize location data. However, while most of these methods model the transition between discrete subsequent locations, they often overlook the broader, global shape of human movements, which remains underexplored in the existing studies. 
As the search for advanced data synthesis methods continues to evolve, it is crucial to explore approaches that outfit the complex data structure of location data. In this regard, Functional Data Analysis (FDA) \citep{RamsaySilverman2005Book, SrivastavaKlassen2016Book} provides a framework for modeling functional datasets, leading to the development of novel algorithms for generating synthetic functional data.

Based on this data modeling, the goal of the research is to develop FDASynthesis, a new algorithm for generating synthetic functional data, where both input and output are functions defined in a suitable functional space. The application case of interest is in the field of human mobility. The proposed methodology is indeed applied to a dataset of Global Positioning System (GPS) trajectories, which describe the spatiotemporal movements of anonymous individuals in a metropolitan area. Each trajectory is modeled as a multidimensional curve that describes the variation of the coordinates and the elapsed time as a function of the record ordering in the trajectory sequence. 
Although designed for the specific case of GPS trajectories, the FDASynthesis procedure is highly scalable and can be applied to the entire range of functional inputs. 

FDASynthesis can be summarized in the following key points. In a preliminary phase, the original raw data are pre-processed and transformed to obtain functional data through smoothing and interpolation techniques. Then, the synthesis of new data occurs iteratively: for each reference original curve, its $k$-closest curves are found and weightedly averaged. This phase is similar to a $k$-nearest neighborhood regression, but in this case, the averaging process uses weights that are stochastically sampled from a probability distribution.
Emphasis is put on choosing the best mathematical embedding for an object-oriented approach. This involves a comprehensive evaluation of the geometric invariance properties of the curves to establish a functional distance metric suited to the specific application. The framework proposed by \citet{SrivastavaKlassen2016Book} has been chosen, as it allows for the analysis of functions by examining their shape, scale, and rotation characteristics \citep{Kurtek2012Statistical}. Besides, by treating amplitude and phase separately, the framework allows isolating and quantifying these aspects of the functions during their comparison and analysis \citep{Tucker2013Generative}.

From a methodological point of view, the research presents for the first time an original object-oriented method for functional data synthesis, introducing significant methodological advancements in SDG for functional data. 
As mentioned above, FDASynthesis works with any functional data. This versatility makes it adaptable to various contexts and extends its usefulness beyond the specific domain of human mobility.
Then, the research is of interest due to its innovative modeling approach, as no prior research on SDG has represented human trajectories as curves within a suitable mathematical space. This novel perspective enhances the analysis and the interpretation of location data and opens new ways of analyzing and understanding human mobility patterns.

The paper is organized as follows. A review of the literature is reported in Section \ref{sec:stateoftheart}, with both a description of the state-of-the-art methodology for data synthesis and a presentation of the evaluation framework. Section \ref{sec:method} is dedicated to explaining the methodology behind FDASynthesis, focusing firstly on the modeling of the trajectory data of the specific case of interest, and then describing the general mathematical framework of analysis and formalizing the algorithm method. The application of the method and its results are described in Section \ref{sec:results}. Finally, Section \ref{sec:conclusions} is dedicated to the discussion of the results and their implication, highlighting possible further development of the research. 

The analysis is carried out by means of the software R \citep{Rsoftware} exploiting some relevant functions from the R package \texttt{fdasrvf} \citep{fdasrvf2024package}. The proposed method has been implemented in R and made available in the \texttt{fdasynthesis} package, which can be found on GitHub in the corresponding repository \citep{fdasynthesis2024package}.

\section{State of the art}\label{sec:stateoftheart}

\subsection{State-of-the-art methods for data synthesis}

The concept of synthetic data as known today was first introduced by \citet{Rubin1993Statistical}.
He suggested using the multiple imputation framework to protect sensitive data by treating unsampled units as missing data and imputing this missing information. The advantage of the procedure is that no original values are included in the released data, offering a high level of protection.
\citeauthor{Rubin1993Statistical}'s proposal was the first to introduce the idea that high-quality synthetic data can serve as a valid substitute for original data \citep{Drechsler2024Thirty}. 
With this new point of view, synthetic data have ceased to be solely an artificial construct for validating methodologies and have become a vector of valuable information in their own right, to be analyzed and exploited without privacy concerns. 
In the past decade, the revolution in the perception of synthetic data has sped up and brought attention back to SDG techniques. First, there has been a growing interest in using large volumes of data to perform analyses and guide data-oriented policies, while concurrently increased concern about the privacy of individuals underlying the data has intensified the generation of new regulations and new privacy-compliant constraints for data collection, publishing, and analysis. 

In this context, in the emerging field of computer science, generative Artificial Intelligence (AI) has started providing synthetic data as a natural output of trained models, assessing their utility by including them in the training process, and suggesting that they can be used as a substitute for original data.
Among these, Generative Adversarial Networks (GANs) are an example used in various application domains \citep{Goodfellow2014Generative}, as well as Variational Auto-Encoders (VAEs) \citep{Kingma2014Auto}. 
Although the domain of generative AI is predominant, the research on SDG methods has also developed on the parallel track of statistical analysis. Unlike the black-box models typical of neural networks, statistical methods inform a model designed for the original data and, by appropriately transforming its input, obtain synthetic data. Examples include random sampling, parametric regression models, and Classification and Regression Trees (CART), as proposed by \citet{Nowak2016Synthpop}, but also Bayesian Network models \citep{Gogoshin2021Synthetic}, Markov models \citep{Lencastre2023Modern}, and Copula-Based models \citep{Datacebo2020}. 
Thus, a primary classification of SDG methods inherits from the research field where their theory is developed. There can be either black-box models of deep-learning approaches, or white-box methods from the statistical field. The method proposed in this research falls into the second group, but in the following there will be reviewed and discussed SDG methods from both categories adhering to this distinction.

As distinct types of data require specifically tailored models, SDG methods
can be classified by the type of data to be synthesized, from standard tabular data to more complex structured data like images and time series. This analysis focuses on multivariate GPS data from smartphones, characterized by groups and sequentiality, i.e., trajectories. Due to this, SDG methods for sequential data are appropriate.
Among deep-learning-oriented methods for sequences, it is to be mentioned the DoppelGANger, a GAN architecture proposed by \citet{Lin2020Using} and implemented on the open-source platform by \citeauthor{Gretel2020Labs} \citep{Lin2020Github, Gretel2020Labs}. The authors start from a Recurrent Neural Network (RNN), a benchmark in the literature for generating synthetic time series, and modify it to overcome its limitations. 
The results of this GAN are remarkable due to the improvements over previous architectures, establishing DoppelGANger as one of the benchmarks in the field.
It is also worth citing the proposal in the Synthetic Data Vault package by DataCebo \citep{Datacebo2020}, that is the Conditional Probabilistic Auto-Regressive (CPAR) model \citep{Zhang2022Sequential}. By combining the principles of autoregressive models with probabilistic NN, it achieves improved accuracy in capturing complex dependencies within sequential data and generating realistic synthetic samples. It is a joint-modelling procedure, aimed at specifying the joint distribution of the data and sampling new data from the fitted probability distribution.  
Other deep-learning methods for sequential data synthesis include TimeGAN \citep{Yoon2019Time}, RCGAN \citep{Esteban2017Real}, and Sig-WGAN \citep{Liao2023Conditional}. In contrast, within the realm of statistical learning, it is important to mention methods that utilize Markov models, proven to outperform AI algorithms in the study by \citet{Lencastre2023Modern}.

The previously mentioned methodologies are general and applicable to any sequential data. However, specific methods have been developed to handle GPS and location data. These methods acknowledge the focus on the mobility application domain and, consequently, the need to develop techniques that account for the unique characteristics of GPS data, such as missing data, latency, variable accuracy, and the recurrence of visited locations.
The first group of works are from the deep-learning domain. \citeauthor{Blanco2022Generation} draw inspiration from the field of Natural Language Processing (NLP) to model GPS trajectories as textual data using a Bidirectional Long Short-Term Memory (BiLSTM) model. The physical space is discretized, and each zone is assigned an identifier so that a trajectory becomes a sequence of letters over time \citep{Blanco2022Generation}.
The PateGail method, proposed by \citet{Wang2023PateGail}, generates virtual trajectories through a Generative Adversary Imitation Learning (GAIL) model, aimed at revealing the underlying patterns in human movements connected to particular places and times. An interesting peculiarity of this architecture is that it removes the risk of exposing personal data by using a method where private data never leaves personal devices: the machinery is not trained on raw GPS data, but only on the synthetic trajectories and a reward measuring the performance of the model. 
The concept of differential privacy is key for the DP-TrajGAN model \citep{Zhang2023Dptrajgan}. The method relies on a GAN that uses Long Short-Term Memory (LSTM) for trajectory generation and suitable differential privacy techniques for balancing privacy and utility for the generative model.

Regarding the statistical domain, the algorithms historically associated with the generation of GPS trajectories are the WHERE model \citep{Isaacman2012Where} and the TimeGeo model \citep{Jiang2016Timegeo}. Both models are probabilistic, aimed at identifying the joint distribution that best describes the spatiotemporal movements of people and sampling synthetic data from it. They use spatial and temporal probability distributions derived from empirical data to infer home, work, and other significant locations for each individual. Then, spatiotemporal patterns are modeled, approximating the individual circadian tendency to travel and people's likely location at different times of the day.   
Expanding on the WHERE model, \citet{Smolak2020Population} add a WHO part, clustering individuals with similar mobility patterns and including the group variability to the model, and a WHEN part, accounting for the temporal patterns and seasonality of human mobility. Similarly, the work by \citet{Pappalardo2018Data} presents the DITRAS method for trajectory generation. It is a two-step methodology that first estimates each user's mobility diary and subsequently integrates the spatial information through the modeling of dynamic Exploration and Preferential Return (d-EPR).
Note that, although the statistical-based models presented are introduced to be used on mobile data, an extension to higher-frequency data such as GPS traces is still possible.
The S$^3$T-Trajectory method, presented by \citet{Zheng2022Utility}, involves generating trajectories using a time-dependent Markov chain. The process is optimized by maximizing three key metrics: data utility, protection from privacy attacks against the individual location, and protection from privacy attacks against the social relationship. 
\citet{Sun2023Synthesizing} propose to integrate the geographic structures into the private trajectory synthesis. They firstly discretize the physical space in cells and estimate the transition probabilities between cells, a-priori forcing to zero the relationship between unreachable cells and thus reflecting the correct mobility traces. Then, a sequence of cells is iteratively sampled, reconstructing the synthetic trajectory dataset.

Among all these methodologies, what is clear is that, despite their differences, they study the spatial-temporal relationships between subsequent observations to determine the mobility trajectory. However, what is missing is a comprehensive view of the trajectory as a whole.
The concept of a global trajectory is present in other application domains, such as path enhancement or trajectory imputation. In these contexts, given a set of GPS points that characterize a trajectory, the goal is to reconstruct the entire path of the individual, filling the gaps between the observed locations. Enhancement methods are also global \citep{Li2021Trajectory}, and it has been already proposed to use spline interpolations to reconstruct the trajectory \citep{Long2016Kinematic}. However, the focus of these works is on extending the trajectory rather than synthesizing new trajectories, which is the primary objective of FDASynthesis.

\subsection{Evaluation framework for SDG methods}

With the exponential increase of SDG algorithms, the need to evaluate their effectiveness in comparison to existing methods is also increasing. It is therefore essential to develop a framework for evaluating synthetic data that provides general criteria for determining their quality in various application contexts. The definition of synthetic data already highlights the two main directions for evaluating SDG methods: privacy and utility. Utility is essential because good synthetic data must be useful in applications, allowing for the same data-driven conclusions that would be reached with the original data. A sufficient condition to ensure this utility is the similarity between the synthetic data and the original data. Synthetic data that is sufficiently similar to the original tends to be more useful, but excessive similarity, approaching a near-replication of the original data, poses a risk to privacy. Privacy requires that the information contained in the original dataset be preserved in such a way that no trace of any individual original data remains, thus making it impossible to identify individuals. This implies a natural trade-off between the utility of synthetic data and its ability to guarantee privacy: the more useful synthetic data is, the greater the risk to privacy, and vice versa. A good framework for evaluating synthetic data must therefore take into account both dimensions, utility and privacy, and highlight potential issues in each. By doing so, it can guide the search for an optimal balance between the utility of synthetic data and the ability to protect the privacy of the original data.

Among utility metrics, the most commonly used are those aimed at measuring the global fidelity of synthetic data to the original data. The most frequent tests include those that assess the equality of the distribution, the correlation, and the autocorrelation of variables. In addition to evaluating variables, key statistical measures such as mean, variance, and range of values are also examined. Visualization tools are often employed to inspect the similarity between two datasets, such as boxplots and histograms for the distribution of continuous variables, as well as correlation matrix plots. In the case of non-tabular data, these analyses are usually replaced by feature analysis. For example, in GPS trajectory data, similarities are measured using features like radius of gyration, travel distance, duration, number of locations visited per trajectory, and mobility entropy. 
Remarkable is the implementation of the open-source Python platform SDMetrics \citep{SDmetrics} that aims at evaluating synthetic tabular data though visual and quantitative checks. Through SDMetric, one can evaluate any tabular synthetic data using pre-implemented metrics dedicated to data quality (comparing variable distributions, variable correlations, variable basic statistics) and data validity (ensuring that no impossible values are present). They also propose a metric for synthetic sequential data, called Multi-Sequence Aggregate Similarity (MSAS), to compare the sequential synthetic data against its real counterpart \citep{Zhang2022Sequential}.
Beyond similarity, other utility metrics assess the the distinguishability between synthetic and original data. Plausibility checks are performed as an a-posteriori control to ensure the plausibility of the obtained values. Moreover, distinguishing games are employed to model the probability of a data point belonging to the synthetic dataset versus the original dataset. Random Forests, Classification Trees, or Logistic Regression models are trained to distinguish the two datasets and, if they never reach good classification performances, they suggest that synthetic data are of good quality. 
More examples on global fidelity utility metrics, also with focus on sequential data and the mobility domain, are presented in the works by \citet{Drechsler2022Challenges, Lin2020Using, Zhang2022Sequential, Dankar2022Multidimensional, Wang2023PateGail}. 

Another category is utility metrics that are analysis-specific. Their goal is to evaluate the similarity between the outputs of analyses conducted on synthetic data compared to those conducted on original data. This approach does not directly check the similarity between the two datasets themselves, but rather focuses on the findings that can be extracted from them. 
For instance, when applying classification and regression methods, one can utilize the Train-on-Synthetic, Test-on-Original (TSTO) metric and the Train-on-Original, Test-on-Synthetic (TOTS) metric. They both involve training a model on one dataset and then evaluating its accuracy on the other dataset. Higher performance indicates better quality synthetic data.
Another approach involves comparing the Train-on-Synthetic, Test-on-Synthetic (TSTS) model with the Train-on-Original, Test-on-Original (TOTO) model. By assessing the differences in the findings between TSTS and TOTO, similar results suggest higher quality synthetic data. This comparison can be done using methods such as ranking algorithms by accuracy, checking for Confidence Interval Overlap, or calculating the Ratio of Estimates. Examples of such analysis, particularly tailored for trajectory data, can be found in the studies by \citet{Bindschaedler2027Plausible, Drechsler2022Challenges, Lin2020Using, Sun2023Synthesizing, Wang2023PateGail}.

Two essential principles guide privacy checks in synthetic data: indistinguishability and uninformativeness \citep{Fiore2020Privacy, Monreale2023Survey}. 
Indistinguishability, proposed by \citet{Sweeney2002Kanonimity}, states that an adversary will not be able to pinpoint an observation to a single specific user. The concept can be generalized for the case of synthetic data as the lack of associations between the original data and the synthetic one. The synthetic data are hence compared to the originals for checking that they are $k$-anonymous, i.e., that synthetic instances are associated with at least $k-1$ other original instances. Ensuring $k$-anonymity typically makes use of techniques such as generalization, microaggregation, and suppression. Generalization involves replacing specific data with broader categories, microaggregation groups data points and replaces them with group averages, and suppression involves removing or aggregating rare and highly identifiable instances. For mobility trajectories, variations of the $k$-anonymity concept have been developed, including $k^{(\epsilon,\tau)}$-anonymity \citep{Gramaglia2017Preserving}, $l$-diversity and $t$-closeness \citep{Tu2018Protecting}.

Uninformativeness ensures that no meaningful or sensitive information about specific individuals can be extracted from the synthetic data, implying that the synthetic data does not leak personal information even when analyzed. In other words, the synthetic data should be resistant to inference attacks that attempt to derive personal information by comparing datasets with and without specific data points.
Building on the foundation of Differential Privacy (DP), new approaches have been proposed, such as $(\epsilon,\delta)$-differential privacy \citep{Shao2013Publishing}, which relaxes the strict requirements of DP to allow for a small probability of failure, and $(k,\gamma)$-plausible deniability \citet{Bindschaedler2027Plausible}, which offers a measure of how plausible it is that an individual's data could have been part of a different dataset. These approaches further enhance privacy by introducing additional flexibility or robustness in the privacy guarantees.

\section{Methodology} \label{sec:method}

\subsection{Functional data representation} \label{subsec:functional}

Before beginning the synthesis procedure, it is necessary to model the input dataset as a functional dataset. This section outlines the preprocessing steps applied in the current application involving GPS trajectories, where the goal is to transform raw GPS signals into continuous functional trajectories. Although the preprocessing described here is specific to the GPS dataset relevant to this particular case study, it can serve as a reference for modeling other types of data in a functional form or be replaced with different pre-processing techniques as needed, depending on the nature of the data being modeled.

The initial dataset consists of GPS measurements that record the positions during the day-to-day displacements of anonimized users at specific times. The fundamental GPS signal is composed of (at least) the two coordinates for the spatial location, $c_1$ and $c_2$, and the absolute time for the temporal indication, $t$, leading to points of the form $P=(c_1, c_2, t)$. The signals describing a single trip of a user are grouped in trajectories. A trajectory describes the unique spatiotemporal movement of a single user through a sequence of GPS data points ordered with respect to the absolute time,  $T_i = \{P_{i1}, \dots, P_{iN_i}\} \quad \forall i=1,\dots, N$. One can hence see the initial dataset as a set of trajectories $\{T_{i}\}_{i=1,\dots,N}$, each composed of sequences of points $P_{ij}=(c_1, c_2, t)_{ij}$ for location and time identification.

Several pre-processing steps are applied to the datasets to make it ready for subsequent functional modeling and analysis. First, since the absolute execution time of the trajectory is not of interest, but rather the timing relative to the starting point, the third component is replaced by the equivalent pair of the start time $t_0$ and the elapsed time since the start, $t_e$. Then, the starting point $t_0$ is ignored, and the analysis is centered solely on $t_e$. 
Next, attention is given to the units of measurement used for the space and the elapsed time. The spatial coordinates are initially converted from the geographic reference system into Universal Transverse Mercator (UTM) coordinates, as the latter uses a simpler Cartesian system in meters and makes distance calculations more straightforward compared to the angular measurements of latitude and longitude. Afterwards, the time and space dimensions are made comparable by normalization. In effect, the discrepancy in units could lead to misleading results if not addressed properly. From the elapsed time measured in seconds and the spatial coordinates in the UTM system in meters, to prevent inaccuracies, all values are normalized between 0 and 1. For example, the first coordinate $c_1$ is transformed to $c_1^\prime$ as:
\begin{equation} \label{eq:normalization}
    c^{\prime}_{1ij} = \frac{\max c_{1\cdot\cdot} - c_{1ij}}{\max c_{1\cdot\cdot} - \min c_{1\cdot\cdot}} \in [0,1] \quad \forall j=1,\dots,N_i \quad \forall i=1,\dots,N
\end{equation}
Consequently, the original dataset is represented by a series of constants: $\min c_{1}$, $\max c_1$, $\min c_2$, $\max c_2$, $\min t_e$ and $\max t_e$, and by the set of trajectories $\{T^\prime_{i}\}_{i=1,\dots,N}$ formed by the points $P^\prime_{ij}=(c_1^\prime, c_2^\prime, t_e^\prime)_{ij}$.

Then, moving to the FDA framework, trajectories are modeled as three-dimensional functions, as expressed in the following Equation \ref{eq:traj_modeling}:
\begin{equation} \label{eq:traj_modeling}
    \begin{split}
        \bm{f}_i & : \, I\rightarrow\mathbb{R}^2 \times \mathbb{R}_{+} \\
        \bm{f}_i & : \, x \mapsto [f_{i}^{(1)}(x), f_{i}^{(2)}(x), f_{i}^{(3)}(x)]
    \end{split}
\end{equation}
They are defined over the interval $I = [0,1]$, where $x \in I$ signifies the percentage of the temporal progression of the trajectory from its beginning to its end. The three dimensions of these functions are used to represent the variations in coordinates ($f^{(1)}$ and $f^{(2)}$) and the elapsed time ($f^{(3)}$), thus fully encapsulating the GPS trajectory information.

To shift from discrete point sequences to their functional representations, various smoothing and interpolation techniques can be employed in the context of FDA. The selection of these techniques should be carefully guided by the specific characteristics of the dataset and the inherent properties of the methods being considered. 
In the current application, spline smoothing techniques are employed for each component separately. For the spatial components $f^{(1)}$ and $f^{(2)}$, natural cubic splines are used, with knots located at the endpoints of the interval and at the measurement points along the trajectory. In contrast, for the time component $f^{(3)}$, which is monotonically increasing, monotone cubic splines are applied with knots positioned according to the measurement sequence.

In conclusion, the preprocessing of the GPS dataset transforms raw positional data into continuous functional trajectories. This transformation yields a dataset of functions, $\{\bm{f}_i\}_{i=1,\dots,N}$, providing a structured foundation for further application to the FDASynthesis methodology.

\subsection{Choice of the appropriate functional metric space} \label{subsec:fdasrvf}

As previously mentioned, the FDASynthesis algorithm involves the search and the averaging of the $K$ closest functions given a reference one. It hence sees the need for an appropriate functional metric space for the analysis that is able to characterize the differences between its elements while respecting the fundamental properties identified for the application of interest. 
Two points are essential: on the one hand, the alignment of the functions and, on the other hand, the attention to the geometric properties of the curves to be preserved.
Generally, registration may be necessary because it provides a measure of variability in amplitude that is not influenced by phase differences, thus improving the interpretability of results and enhancing the statistical power of the analyses. In the specific case of the current research, the registration enables considering variations in the sampling frequency of GPS data, which might lead to phase misalignment. These are the reasons why the proposed approach involves decomposing the distance into two dimensions: amplitude distance and phase distance. The amplitude metric assesses the similarity of the aligned functions, while the phase metric evaluates the effort required to align the functions. Then, it is desirable that the space used is as flexible as possible, considering the analysed functions in all their features of interest and in particular preserving their geometrical invariance properties. For example, choosing the $L^2$ distance between curves would be suitable only when measuring the difference in amplitude of the functions, but this is not always the case. Functional data samples in various fields show properties of rotation invariance and/or scale invariance (e.g., when data represent boundary shapes, or for signature data, as in \citet{SrivastavaKlassen2016Book}) that the functional space must maintain to ensure consistent analytical results.

For these reasons, the chosen modeling framework is functional shape analysis, as proposed in the pioneering work by \citet{Tucker2013Generative}. The authors suggest tools for functional statistical analysis exploiting the definition of the Square-Root Velocity Function (SRVF) and the Fisher–Rao Metric. Their work has been further extended in the book by \citet{SrivastavaKlassen2016Book}, presenting a comprensive framework of curve shape analysis where rotational and scale invariance is also taken into account. In fact, the shape analysis framework is flexible and allows working with various scenarios, combining the properties of shape, size, and orientation to use different representations for functional spaces \citep{Kurtek2012Statistical}. For the case of interest, functions without rotational invariance and without scale invariance are employed, as in \citet{Tucker2013Generative}, since they are suitable for the application involving mobility curves. However, this does not constrain the proposed methodology, which can be easily adapted to incorporate invariances into the SDG process with a straightforward theoretical and algorithmic modifications.
For a comprehensive examination, please refer to the seminal works by \citet{Kurtek2012Statistical}, \citet{Tucker2013Generative}, and \citet{SrivastavaKlassen2016Book}. The following overview outlines the fundamental concepts for a thorough understanding of the proposed methodology and the applications of the framework discussed. Note that in what follows, the norm $\lVert\cdot\rVert$ refers to the norm $L^2$ if no subscript is specified, while when a subscript is provided, it denotes a norm associated with the different corresponding space.

Let $I\subset \mathbb{R}$ be any interval domain. As every interval can be mapped into the unitary one, for the sake of simplicity, let $I=[0,1]$.
Consider the space of $p$-dimensional absolutely continuous functions over $I$:
\begin{equation*}
    \mathcal{F}(I)=\{\bm{f} : I\rightarrow \mathbb{R}^p \in \text{AC}(I)\}
\end{equation*}
Then, given any function \(\bm{f} \in \mathcal{F}(I)\), its Square-Root Velocity Function (SRVF) is defined as the $p$-dimensional function $\bm{q}:I\rightarrow \mathbb{R}^p,\: \bm{q}\in L^2(I),$ given by:
\begin{equation} \label{eq:srvf}
    \bm{q}(x) = 
    \begin{cases}
        \frac {\bm{\dot{f}}(x)}{ \sqrt{\lVert\bm{\dot{f}}(x)\rVert}}  & \text{if } \bm{\dot{f}}(x) \neq \mathbf{0} \\
        \mathbf{0} & \text{otherwise}\\
    \end{cases}
\end{equation}
The correspondence between a function and its SRVF is bijective up to translations, such that for every \(\bm{q} \in L^2(I)\), there exists a unique class of curves equivalent up to translations for which \(\bm{q}\) is its SRVF. It is indeed easy to prove that:
\begin{equation}\label{eq:srvfm1}
    \bm{f}(x) = \bm{f}(0) + \int_0^x \bm{q}(t) \lVert\bm{q}(t)\rVert dt \quad \forall x \in I 
\end{equation}
This property implies that, if the analysis is based on the SRVFs, any distance will be systematically made invariant with respect to the starting position.

To define amplitude and phase distance, it is essential to introduce the concept of reparametrization through the action of warping functions. Hence, consider the set of all warping functions on $I$:
\begin{equation*}
    \Gamma = \{\gamma: I \rightarrow I \quad | \quad \gamma (0) = 0; \gamma (1)=1; \gamma \text{ is a diffeomorphism}\}
\end{equation*}
The action of a warping function $\gamma \in \Gamma$ on a function $\bm{f} \in \mathcal{F}(I)$ induces a corresponding action on its respective SRVF $\bm{g} \in L^2(I)$. Specifically, if a function $\bm{f}$ is warped by $\gamma$, the SRVF of $(\bm{f} \circ \gamma)$ is $(\bm{q} \circ \gamma) \sqrt{\dot{\gamma}}$. Starting from this expression, it can be proven that the distance between the SRVFs of aligned functions is isometric under warping. Indeed, $\lVert (\bm{q}_1 \circ \gamma)\sqrt{\dot{\gamma}} - (\bm{q}_2 \circ \gamma)\sqrt{\dot{\gamma}} \rVert = \lVert \bm{q}_1 - \bm{q}_2 \rVert$, differently from the $L^2$ norm between functions where $\lVert \bm{f}_1 \circ \gamma - \bm{f}_2 \circ \gamma \rVert \neq \lVert \bm{f}_1 - \bm{f}_2 \rVert$. This is what suggests the definition of the so-called amplitude distance.
The amplitude distance between two functions $\bm{f}_i, \bm{f}_j \in \mathcal{F}(I)$ is defined as the minimum $L^2$ distance between SRVFs of the optimally-aligned functions, that is:
\begin{equation} \label{eq:d_a}
    D_a(\bm{f}_i, \bm{f}_j) = \min_{\gamma \in \Gamma} \: \lVert \; \bm{q}_i - (\bm{q}_j \circ \gamma) \sqrt{\dot{\gamma}} \; \rVert
\end{equation}
It is a proper distance in the quotient space $\mathcal{F}(I)/\Gamma$ since it is symmetric, satisfies the triangle inequality, and $D_a(f_i, f_j) = 0$ implies that $f_i$ and $f_j$ are in the same equivalence class, i.e., equal up to reparametrization in $\Gamma$. This makes $D_a$ a semi-distance in the space of the original functions $\mathcal{F}(I)$.
Furthermore, when the $L^2$ metric is applied specifically to SRVFs of functions whose first derivative is strictly positive, like cumulative distribution functions, it matches exactly with the well-known Fisher-Rao Riemannian metric.

Given the amplitude distance, the notion of mean function comes straightforward. The mean function of a dataset of curves $\{\bm{f}_i\}_i\subset\mathcal{F}(I)$ can be defined as:
\begin{equation} \label{eq:meanf}
    \bm{\mu}_f = \argmin_{\bm{f}\in\mathcal{F}(I)} \sum_i D_a(\bm{f}, \bm{f}_i)^2
\end{equation}
The $\bm{\mu}_f$ function is called Karcher mean, or Frechet mean. An alternative definition of the mean is possible when working in the $L^2(I)$ space of SRVFs, and not in the $\mathcal{F}(I)$ space of the original curves. Indeed, the function $\bm{\mu}_q$, defined below, is the SRVF of the function $\bm{\mu}_f$:
\begin{equation} \label{eq:meanq}
    \bm{\mu}_q = \argmin_{\bm{q}\in L^2(I)} \sum_i \min_{\gamma_i\in\Gamma}\lVert \bm{q}- (\bm{q}_i\circ\gamma_i)\sqrt{\dot{\gamma_i}}\rVert ^2
\end{equation}
Finally, starting from Definitions \ref{eq:meanf} and \ref{eq:meanq}, one can easily define the weighted version of the average functions. For instance, given a set of weights for each function $\{p_i\}_i$ satisfying $p_i \in [0,1] \:\forall i$ and $\sum p_i = 1$, the weighted mean function in the SRVF space is:
\begin{equation} \label{eq:meanq_2}
    \bm{\mu}_q = \argmin_{\bm{q}\in L^2(I)} \sum_i p_i \, \min_{\gamma_i\in\Gamma}\lVert \bm{q}- (\bm{q}_i\circ\gamma_i)\sqrt{\dot{\gamma_i}}\rVert ^2
\end{equation}

For the elements $\gamma\in\Gamma$, it is also possible to define the corresponding SRVF as in Definition \ref{eq:srvf} that, given the properties of warpings, can be simplified as:
\begin{equation}
    \psi(x) = \sqrt{\dot{\gamma}(x)} \quad \forall x \in I
\end{equation}
Moreover, it is straightforward to prove that the SRVF of any warping lies in the Hilbert sphere, i.e., the space of $L^2$ functions with unitary norm:
\begin{equation*}
    S_1^\infty (I) = \{\psi \in L^2(I) \quad |\quad \lVert \psi\rVert = 1\}
\end{equation*}
Now, let $\gamma_{opt} \in \Gamma$ be the minimizer of the quantity in Definition \ref{eq:d_a} and $\psi_{opt} \in S_1^\infty(I)$ its SRVF. Consider the identity warping and its SRVF $\psi_0\in S_1^\infty(I)$, that is, the constant function $\psi_0\!:\!x \mapsto 1$. Then, the phase distance between the two functions $\bm{f}_i$ and $\bm{f}_j$ can be defined as the distance between $\psi_{opt}$ and $\psi_0$ computed on the sphere $S_1^\infty (I)$, that is:

\begin{equation}\label{eq:d_p}
    D_p(\bm{f}_i, \bm{f}_j) = \lVert \; \psi_0 - \psi_{opt} \; \rVert _{S_1^{\infty}(I) } = \cos ^{-1}\left(\int_0^1 \psi_0(x) \psi_{opt}(x) dx\right)
\end{equation}

\subsection{Proposed synthetic data generation algorithm} \label{subsec:fdasynthesis}

The FDASynthesis method is implemented as a two-step approach, where first distances between curves are computed, and then the proper synthesis phase occurs. An outline of the methodology is presented in Algorithm \ref{scheme:algo}.

\begin{algorithm}[tb]
    \KwData{Original functional dataset $\{\bm{f}_i\}_{i=1,\dots,N} \subset \mathcal{F}(I)$}
    \KwIn{$\delta\in [0,1]$; $K\in\mathbb{N}$; $\alpha_0 \in [1,\infty)$}
    \BlankLine
    \tcc{Computing distances between functions:}
    \For{$i, j \in \{1, \dots, N\}; i \neq j$}{
        $D_a(\bm{f}_i, \bm{f}_j)$ as in Eq. \ref{eq:d_a} \tcp{amplitude distance}
        $D_p(\bm{f}_i, \bm{f}_j) $ as in Eq. \ref{eq:d_p} \tcp{phase distance}
        $\textbf{D}[i,j] = D(\bm{f}_i, \bm{f}_j) = \delta D_a(\bm{f}_i, \bm{f}_j) + (1-\delta) D_p(\bm{f}_i, \bm{f}_j)$ \tcp{global distance matrix}
    }
        
    \tcc{Iteratively synthesize new data:}
    \For{$i \in \{1, \dots, N\}$}{
        \tcc{Find the K closest curves:}
        $\bm{d}_i = [d_{i1}, \dots, d_{iK}]$ with $d_{ik} = sort(\textbf{D}[i,:])[k+1]$ \tcp{distances to the K closest curves}
        \tcc{Compute their weights:}
        $\bm{\alpha}_i = [\alpha_{i1}, \dots, \alpha_{iK}]$ where $\alpha_{ik} = \alpha_0 \tfrac{e^{-d_{ik}}}{\sum_{n=1}^K e^{-d_{in}}} $ \tcp{Dirichlet parameters}
        $\bm{p}_i=[p_{i1}, \dots, p_{iK}] \sim Dir(\bm{\alpha}_i)$ \tcp{weights for the average}
        \tcc{Perform the weighted average:}
        $\bm{q}_i^s$ as in Eq. \ref{eq:meanq_3} \tcp{average SRVF}
        $\bm{f}_{i0}^s$ as in Eq \ref{eq:initial_pts} \tcp{average initial point}
        $ \bm{f}_i^s(x) = \bm{f}_{i0}^s + \int_0^x \bm{q}_i^s(t) \lVert\bm{q}_i^s(t)\rVert dt$ as in Eq. \ref{eq:mean_fsynth}
    }
    \BlankLine
    \KwResult{Synthetic functional dataset $\{\bm{f}_i^s\}_{i=1,\dots,N} \subset \mathcal{F}(I)$}
    
    \caption{Pseudo-algorithm of the FDASynthesis SDG method. The distance computation is summarized in lines 1-5, and the data synthesis in lines 6-13.}
     \label{scheme:algo}
\end{algorithm}

The amplitude and phase separation framework leaves two notions of distance for functions: one related to amplitude and the other to phase, as defined, respectively, in Equations \ref{eq:d_a} and \ref{eq:d_p}. For this work, a single measure of dissimilarity between functions is of interest and would be even more effective if it accounted for both sources of dissimilarity.
To address this, a new distance function is defined to jointly consider phase variability and amplitude variability. 
This combined distance function, referred to as $D$, is defined between every pair of functions $\bm{f}_i, \bm{f}_j \in \mathcal{F}(I)$ as the convex combination of their amplitude distance $D_a$ and their phase distance $D_p$:
\begin{equation} \label{eq:Dtot}
    D(\bm{f}_i, \bm{f}_j) = \delta D_a(\bm{f}_i, \bm{f}_j) + (1-\delta) D_p(\bm{f}_i, \bm{f}_j)
\end{equation}

The parameter $\delta\in[0,1]$ plays a crucial role in this process, indicating the relative importance assigned to amplitude and phase in the overall distance measure. The details of the tuning procedure for this parameter are discussed in the next Subsection \ref{subsec:tuning}.
Once the combination parameter $\delta$ is chosen, the distances are calculated as in Equation \ref{eq:Dtot} for all pairs of functions in the dataset $\{\bm{f}_i\}_{i=1,\dots,N} \subset \mathcal{F}(I)$. All these values are then stored in a matrix $\textbf{D}\in \mathcal{M}_{N\times N}(\mathbb{R})$ of elements $\textbf{D}[i,j] = D(\bm{f}_i, \bm{f}_j) \;\; \forall i, j \in \{1, \dots, N\}$. 
This procedure is computationally expensive, as it requires calculating $N(N-1)/2$ optimal alignments between pairs of functions and their corresponding distances. However, this computation needs to be performed only once at the beginning. Additionally, the implementation is designed to run in parallel across multiple cores, significantly reducing the overall computation time.

The core data generation technique starts after the computation of the distance matrix $\textbf{D}$. The synthesis operates on a function-by-function basis, generating one synthetic trajectory for each original one. First, given a reference function from which to create a corresponding synthetic twin, it carefully selects the $K$ instances that have the shortest combination of amplitude and phase distance from the reference function. This selection ensures that the generated trajectories closely resemble the original functions in terms of their dynamic characteristics. Then, when these instances are identified, their SRVFs are used to perform a weighted average and the results are mapped back from the SRVF space to the original functional space. The averaging process effectively captures the essential features of the curves while smoothing out the variations of highly identifiable outlier patterns. By mitigating the influence of individual outliers, the process ensures that specific, potentially sensitive information from any single trajectory is less likely to be inferred from the generated data.
The approach resembles the $k$-nearest neighbors ($k$-nn) algorithm because it involves identifying and averaging the most similar instances. However, this method introduces a stochastic element to the weighted averaging process by sampling weights from a Dirichlet distribution. The use of a random, and non-deterministic, procedure aims at enhancing the privacy of the users in the original dataset since it introduces variability that can obscure the exact influence of any single data point. By sampling weights from a Dirichlet distribution, the method ensures that the final prediction is less directly tied to individual training examples, thus reducing the risk of inferring sensitive information about any particular user. This stochastic approach acts as a form of differential privacy, where the randomness in the weight assignment helps to protect the privacy of the data used in the averaging process, making it more challenging to reverse-engineer the original data from the predictions.

Formally, given a dataset of $N$ functions $\{\bm{f}_i\}_{i=1,\dots,N} \subset \mathcal{F}(I)$, the reference function of interest $\bm{f}_i$ is chosen and its $K$ closest functions $\{\bm{f}_{k|i}\}_{k=1,\dots,K}$ are determined. Note that, in what follows, the index $i=1,\dots,N$ refers to objects associated to the fixed reference function, while the indices $\{k|i\}_{k=1,\dots,K}$ represent the $K$ indices of the neighbor functions ordered with respect to their distance from the reference function $i$.
Then, let $\bm{d}_i = [d_{i1}, \dots, d_{iK}]$ represents the $K$-dimensional vector of distances between the reference function $i$ and its $K$ nearest functions, whose elements are denoted as:
\begin{equation}
    d_{ik} = \mathcal{D}(\bm{f}_i, \bm{f}_{k|i}) \quad \forall k=1,\dots,K
\end{equation}

First, define the parameter vector $\bm{\alpha}_i=[\alpha_{ik},\dots,\alpha_{iK}]$ whose elements are computed as transformation of the distances as follows:
\begin{equation} \label{eq:alphas}
    \alpha_{ik} = \alpha_0 \tfrac{g(d_{ik})}{\sum_{n=1}^K g(d_{in})} \quad \forall k=1,\dots,K; \quad \sum_{k=1}^K \alpha_{ik} = \alpha_0 >0
\end{equation}
The function $g: \mathbb{R}^{+} \rightarrow \mathbb{R}^{+}$ in Equation \ref{eq:alphas} can be any monotone non-increasing function which describes an inverse relationship between the weights and the distances. In this study, it is chosen to be $g(x) = e^{-x}$, but possible alternatives are the equilateral hyperbola $g(x)=\tfrac{1}{1+x}$ and parametrized variations such as $g(x) = e^{-\beta_0 x} \text{ with } \beta_0 > 0$. The parameter $\alpha_0 > 0$ controls the sum of the concentration parameters, and its tuning will be discussed later in this Section.

The weight vector $\bm{p}_i = [p_{i1},\dots,p_{iK}]$ is then randomly sampled from the Dirichlet distribution with concentration parameters equals to the parameter vector $\bm{\alpha}_i$:
\begin{equation}
    \bm{p}_i \sim Dir (\bm{\alpha}_i)
\end{equation}

In line with the choice of reference space for the functions, the weighted average is computed as the weighted mean of the SRVF of the $K$ neighbor curves and then mapped back to the original space with the correct translation of the starting point, as expressed in Equation \ref{eq:meanq_2} and Equation \ref{eq:srvfm1}. Given the SRVF transformations of the neighbors $\{\bm{q}_{k|i}\}_{k=1,\dots,K}$, the average SRVF is computed precisely as:
\begin{equation} \label{eq:meanq_3}
    \bm{q}_i ^s = \argmin_{\bm{q}\in L^2(I)} \sum_{k=1}^K p_k  \min_{\gamma_k\in\Gamma}\lVert \bm{q}- (\bm{q}_{k|i}\circ\gamma_k)\sqrt{\dot{\gamma_k}}\rVert ^2
\end{equation}
The mean starting point is determined as the arithmetic weighted mean of the starting points: 
\begin{equation} \label{eq:initial_pts}
    \bm{f}_{i0}^s  \: = \: \sum_{k=1}^K p_k \bm{f}_{k|i}(0)
\end{equation}
Combining the mean SRVF of Equation \ref{eq:meanq_3} and the mean starting point of Equation \ref{eq:initial_pts}, the mean function in the original space is found through the inverse mapping:
\begin{equation} \label{eq:mean_fsynth}
    \bm{f}_i^s(x) = \bm{f}_{i0}^s + \int_0^x \bm{q}_i^s(t) \lVert\bm{q}_i^s(t)\rVert dt \quad \forall x \in I
\end{equation}
The mean function $\bm{f}_i^s$ computed at each iteration $i$ serves as the synthetic datum generated during that iteration. The entire set of these $N$ mean functions forms the output synthetic dataset $\{\bm{f}_i^s\}_{i=1,\dots,N} \subset \mathcal{F}(I)$.

\subsection{Hyperparameter tuning} \label{subsec:tuning}

What remains to be discussed is the tuning procedure of the three input parameters $\delta$, $K$, and $\alpha_0$.
The choice of the parameter $\delta\in[0,1]$ is crucial for assigning the proper importance weight to the two distances involved in the procedure. The extreme cases of $\delta=1$ and $\delta=0$ would lead to the use of, respectively, only amplitude distance and only phase distance. A value of $\delta\in(0,1)$ would contrarily average the information of both distances and obtain the final distance as a convex combination of the original one. 
This parameter can be chosen in advance, based on the prior knowledge of the specific application or interest. Alternatively, $\delta$ can be determined through parameter selection methods, allowing for a more empirical approach to balance the contributions of amplitude and phase in the distance calculation. This study however exploits the data-driven approach of \citet{Bellanger2021Methode} to gain insights into the significance of the parameter $\delta$. Precise details on the method are not discussed here, but can be found in the original paper by \citet{Bellanger2021Methode} and in the documentation of its package by \citet{spartaas2023package}. In summary, the method works by averaging two sources of information into one as the best compromise with respect to a defined loss function to be minimized. This criterion is based on the cophenetic correlation matrix and works iteratively to find the value of $\delta$ for which the difference in the correlation of the cophenetic matrix of $D$ with $D_a$ and the cophenetic matrix of $D$ with $D_p$ is minimized. By opting for this parameter selection method, the goal is first to explore whether a compromise can effectively merge the two sources of distance and, if so, to determine the optimal balance. Solutions where no unique optimal compromise is present can be easily identified through the analysis of the shape of cophenetic correlation difference curve when varying $\delta$. Flat curves symbolize either similar sources and redundancy of the information, leading to the choice of any $\delta \in[0,1]$ without loss of information, or missing optimal compromise, forcing the use of only one source of information with $\delta \in \{0,1\}$. Contrarily, if an optimal compromise exists, the parameter will be selected as the one that minimizes the discrepancy of the cophenetic distances and best achieves this compromise, $\delta \in (0,1)$.

The choice of the two input parameters $K$ and $\alpha_0$ remains to be discussed. $K\in \mathbb{N}$ represents the neighborhood size, affecting performance as in a standard $k$-nn approach: larger $K$ values result in a greater smoothing effect, while smaller $K$ values generate peculiar elements that resemble the existing ones. 
A similar principle applies to the parameter $\alpha_0\in[1,\infty)$, which controls the sum of the concentration parameters in the Dirichlet distribution. $\alpha_0$ is chosen to be greater than $1$ to prevent the distribution from concentrating on the extremes of the simplex. $\alpha_0$ values close to $1$ yield a uniform distribution over the simplex, leading to the sampling of components that are equally probable, while higher values concentrate probability on components with larger $\alpha_k$. High $K$ and low $\alpha_0$ therefore result in an averaging effect on the components, while low $K$ and high $\alpha_0$ do the opposite. Proper tuning is necessary to mitigate the averaging effect and thus maintain the global characteristics of the dataset while preserving the non-identifiability of the synthetic data. 

This study proposes a method for tuning these two parameters, aimed at finding the right balance between the utility of the synthetic output dataset and the privacy compliance. The procedure is divided into two consecutive phases. 
The first phase aims at finding an optimal value of $\alpha_0$ for each $K$ with focus on privacy concerns. For different combinations of $\alpha_0$ and $K$, synthetic data are generated and the distances between the synthetic and the original data are measured. The minimum value of these distances is taken as a measure of privacy preservation, as low distances indicate high similarity between the synthetic and original data, and therefore a higher risk of privacy disclosure. The goal here is to exclude a-priori those parameter combinations that do not preserve privacy by looking at the minimum distances and selecting the optimal combinations. At the end of the first phase, an optimal value of $\alpha_0$ is selected for each value of $K$, obtaining $\hat{\alpha}_0(K)$. To also determine the optimal value of $K$, the second phase evaluates the similarity between the distances among synthetic curves and the distances among real curves. In this phase, the goal is to ensure the quality of the synthetic data in terms of overall similarity to the original data. In practice, for various values of $K$, and keeping the optimal value of $\hat{\alpha}_0(K)$ fixed, the two distance matrices are computed and their correlation is analyzed. The value of $K$ that maximizes this correlation is preferred, which finally leads to the choice of the optimal combination of parameters, $\hat{K}$ and $\hat{\alpha}_0 = \hat{\alpha}_0(\hat{K})$.

To mitigate the high computational cost of this tuning procedure, a simplified version is proposed. 
The authors suggest to use a clustering algorithm to assign each original curve to a cluster and then to measure exclusively the relationships between curves within the same cluster. Rather than examining the distance matrices between all synthetic and original curves, only relationhips betwwen curves within the same cluster are investigated. This results in a great decrease in the computational complexity of the tuning method. The total number of distance calculations is reduced to $\mathcal{O}(\sum_{g=1}^G N_g^2)$, where $N_g$ is the number of curves in the $g$-th cluster and $G$ is the number of clusters, which is significantly smaller than the $\mathcal{O}(N^2)$ of the original tuning procedure. Moreover, despite the reduction in computational effort, this approach remains effective because the clustering process ensures that the relationships between similar curves, which are more likely to be informative for tuning the parameters, are still accurately captured. 
This strategy will now be formalized and examined in detail. A scheme of the two tuning phases is also outlined in Algorithms \ref{scheme:tuning1} and \ref{scheme:tuning2}.

\begin{algorithm}[tb] 
    \KwData{Original and synthetic functional datasets with grouping structure, $\{\bm{f}_{ig}\}_{i=1,\dots,N_g; g=1,\dots,G}$ and $\{\bm{f}_{ig}^s\}_{i=1,\dots,N_g; g=1,\dots,G}$}
    \KwIn{$K$ values: $\{K_{t_k}\}_{t_k=1,\dots,N_k}$; $\alpha_0$ values: $\{\alpha_{0t_{\alpha}}\}_{t_{\alpha}=1,\dots,N_{\alpha}}$ }
    \BlankLine
    \For{$t_k = 1,\dots, N_k$}{
        init $\bm{I}^{(1)}$ of length $N_{\alpha}$ \tcp{initial empty vector of minima}
        \For{$t_{\alpha} = 1, \dots, N_{\alpha}$}{
            \For{$g=1,\dots,G$}{
                $\textbf{D}_g^{(1)}$ as in Eq. \ref{eq:tuning:D1} \tcp{distances between synthetic and original}
            }
            $\bm{I}^{(1)}[t_{\alpha}] = \min_{g} \min \textbf{D}_g^{(1)}$ as in Eq. \ref{eq:tuning:i1} \tcp{index: minimum value}
        } 
        choose $\hat{\alpha}_0=\hat{\alpha}_0(K)$ for which $\bm{I}^{(1)}[t_{\hat{\alpha}}]$ stabilizes \tcp{criterion 1}
        or choose $\hat{\alpha}_0=\hat{\alpha}_0(K)$ as the first for which a threshold is overcome \tcp{criterion 2}
    }
    \BlankLine
    \KwResult{Optimal $\alpha_0$ values $\{\hat{\alpha}_0(K_{t_k})\}_{t_k=1,\dots,N_k}$}
    \caption{Pseudo-algorithm of the first tuning phase. Here it is used the stabilization criterion (line 9), although also the first minimum-above-threshold could be used (line 10).}
    \label{scheme:tuning1}
\end{algorithm}

\begin{algorithm}[tb]
    \KwData{Original and synthetic functional datasets with grouping structure, $\{\bm{f}_{ig}\}_{i=1,\dots,N_g; g=1,\dots,G}$ and $\{\bm{f}_{ig}^s\}_{i=1,\dots,N_g; g=1,\dots,G}$}
    \KwIn{Combinations of $K$ and optimal $\alpha_0(K)$ values: $\{(K, \hat{\alpha}_0(K))_{t_k}\}_{t_k=1,\dots,N_k}$ }
    \BlankLine
    init $\bm{I}^{(2)}$ of length $N_k$ \tcp{initial empty vector of correlations}
    \For{$t_k = 1,\dots, N_k$}{
        init $\bm{\rho}$ of length $G$ \tcp{initial empty vector of correlations}
        \For{$g=1,\dots,G$}{
            $\textbf{D}_g^{(2)}$ as in Eq. \ref{eq:tuning:D2} \tcp{distances between original}
            $\textbf{D}_g^{s(2)}$ as in Eq. \ref{eq:tuning:D2} \tcp{distances between synthetic}
            $\rho_g = cor (\textbf{D}_g^{s(2)}, \textbf{D}_g^{(2)}) $ as in Eq. \ref{eq:tuning:D2} \tcp{within-group correlation}
        }
        $\bm{I}^{(2)}[t_{k}] = \text{P}_{(25)} \{ |\rho_g| \}$ as in Eq. \ref{eq:tuning:i2} \tcp{indicator: $25$-th percentile}
    }
    choose $\hat{K} = K_{t_k} = \argmax_{t_k} \bm{I}^{(2)}[t_{k}]$ \\
    set $\hat{\alpha}_0 = \hat{\alpha}_0(\hat{K})$
    \BlankLine
    \KwResult{Optimal combination $(\hat{K}, \hat{\alpha}_0)$}
    \caption{Pseudo-algorithm of the second tuning phase.}
    \label{scheme:tuning2}
\end{algorithm}

The original functions grouping exploits the hierarchical clustering method using the distance $\textbf{D}$ and the complete linkage. The dendrogram is then partitioned using a dynamic cut with adaptive threshold on the distance, as proposed by \citet{Langfelder2008Defining}, getting clusters with certain properties on size, density and splitting tendency (i.e., getting small or large clusters). This approach provides a more flexible and robust solution compared to a fixed-cut hierarchical clustering for situations involving complex data structures like those under analysis. However, the authors emphasize that any clustering method can be employed depending on the characteristics of the data, with hierarchical clustering and dynamic cuts being just one option among many.

The original dataset of functions, integrated with the grouping structure, is then written as:
\begin{equation}
    \{\bm{f}_{ig}\}_{i=1,\dots,N_g; g=1,\dots,G}
\end{equation}
where curve $\bm{f}_{ig}$ is the $i$-th curve belonging to the cluster $g$-th, given the total number of curves in cluster $g$, $N_g$, and the total number of clusters, $G$.
 
The group labels are then associated with the synthetic functions, resambling the correspondence between original and synthetic functions of the FDASynthesis procedure. The method produces one synthetic curve for each original curve, each synthetic curve is assigned to the same cluster as its corresponding original curve:
\begin{equation}
    \{\bm{f}_{ig}^s\}_{i=1,\dots,N_g; g=1,\dots,G}
\end{equation}
where curve $\bm{f}^s_{ig}$ is the synthetic curve associated with $\bm{f}_{ig}$, the $i$-th curve belonging to the $g$-th cluster.

In the first tuning phase, the distances between synthetic and real curves belonging to the same clusters are considered. Formally, $G$ matrices $\textbf{D}_g^{(1)} \in \mathbb{M}^{N_g \times N_g}$ are constructed as described in the following expression:
\begin{equation}\label{eq:tuning:D1}
     \textbf{D}_g^{(1)} \text{ with } \textbf{D}_g^{(1)}[i,j] = D(\bm{f}^s_{ig}, \bm{f}_{jg}) \quad \forall i,j=1,\dots,N_g \quad \forall g=1,\dots,G
\end{equation}
Among these distance values, the minimum if the minima is chosen as an indicator for the tuning methodology. This minimum distance indicator can be written as follows:
\begin{equation} \label{eq:tuning:i1}
    I^{(1)} = \min_{g=1,\dots,G} \min \textbf{D}_g^{(1)}
\end{equation}
The vector of indicator $\bm{I}^{(1)}$ is built, collecting the minimum distance indicators when varying $\alpha_0$ for each $K$. Then, an optimal $\alpha_0$ is chosen for each $K$ according to a privacy-preservant criterion on the vector of indicators $\bm{I}^{(1)}$. Optimality can be assessed by observing the stabilization of the curves describing the minimum distance indicator when $\alpha_0$ increases at fixed $K$. In fact, such curves form an elbow, after which the minimum distance continues to increase but at a reduced intensity. Alternatively, another criterion involves establishing a minimum privacy risk threshold and evaluating the first $\alpha_0$ value for which such threshold is exceeded. While the elbow criterion is data driven, this second rule is lead by the choice of the threshold that might be suggested by the distribution of the original distances, or fixed a priori to ensure a certain level of privacy protection. 

In the second phase, the distances between curves belonging to the same cluster are examined both for the original and the synthetic dataset. The correlation between the distance matrices is then calculated for each cluster, resulting in $G$ correlation values, $\rho_g$. 
The matrices and their correlation can be formulated as follows:
\begin{equation}\label{eq:tuning:D2}\begin{split}
    \textbf{D}_g^{s(2)}  \text{ with } \textbf{D}_g^{s(2)}[i,j]&= D(\bm{f}^s_{ig}, \bm{f}^s_{jg}) \quad \forall i,j=1,\dots,N_g \quad \forall g=1,\dots,G \\
    \textbf{D}_g^{(2)}  \text{ with } \textbf{D}_g^{(2)}[i,j]&= D(\bm{f}_{ig}, \bm{f}_{jg}) \quad \forall i,j=1,\dots,N_g \quad \forall g=1,\dots,G \\
    \rho_g &= cor (\textbf{D}_g^{s(2)}, \textbf{D}_g^{(2)}) \quad \forall g=1,\dots,G
\end{split}\end{equation}
The $25$-th percentile is used as a representative indicator of the $G$ absolute correlations $|\rho_g|$. Opting for the $25$-th percentile instead of the mean ensures that at least $25$\% of the clusters exhibit a high correlation, offering a more robust measure. In contrast to selecting the minimum, using the $25$-th percentile helps mitigate the impact of potential outliers in the distribution, making the indicator more reliable. Therefore, the indicator is defined as follows:
\begin{equation}\label{eq:tuning:i2}
    {I}^{(2)} = \text{P}_{(25)} \{ |\rho_g| \}
\end{equation}
Then, the value of $K$ which maximizes the vector of indicators $\bm{I}^{(2)}$, $\hat{K}$, and the corresponding $\hat{\alpha}_0 = \hat{\alpha}_0(\hat{K})$ are chosen as the optimal combination of parameters and are used in the FDASynthetis method.

\section{Application and results}\label{sec:results}

\subsection{Study case: GPS data overview}

The FDASynthesis method is suitable for application to any type of functional data. As mentioned, this study specifically tested it on a dataset of real smartphone GPS data.
The dataset of GPS signals is provided by Cuebiq Inc. through the ``Data for Good" program. Cuebiq, a location intelligence firm, collects and analyzes anonymized location data with strong privacy measures, including GDPR compliance and advanced anonymization techniques. Their ``Data for Good" initiative offers privacy-preserving data for academic research on human mobility\footnote{The authors adhered to their licensing agreement and did not attempt to re-identify anonymized users or their trajectories.}.
The data set provided includes anonymized smartphone geolocalization records detailing location information at various times for Cuebiq users, identified only by anonymized IDs. It also provides additional details, such as GPS measurement accuracy and the type of smartphone operating system used for data collection.

The trajectory observations analysed in this study are located in the Milan municipality and its surroundings and refer to the period from December 1st to December 20th 2019 during working days (from Monday to Friday) and daily hours (from 7:30 in the morning to 7:30 in the evening). Trajectories are selected with a common starting point, namely, the Bovisa campus of Politecnico di Milano, with the goal of highlighting the mobility patterns of users living in the metropolitan area and commuting from the specific point of interest. 
Other filters have been applied to ensure the good quality of the location data. The final dataset retains only trajectories with sufficient coverage (more than $5$ distinct location points), with limited gaps between consecutive records both in space ($\leq$3km) and in time ($\leq$30min), sufficiently high accuracy ($\leq$1200m), and plausible values for mobility measures in the urban context (point speed smaller than 90km/h). 
The final dataset is composed of $N=1,216$ trajectories of $745$ distinct anonymized users for a total of $11,928$ GPS signals. Additional statistics of the main features of the analyzed dataset are reported in Table \ref{tab:GPS_stats}.
These GPS data are extracted, then transformed and modeled as functions before being subjected to the FDASynthesis method, as detailed in Subsection \ref{subsec:functional}, ending up with a dataset $\{\bm{f}_i\}_i$ of $N=1,216$ three-dimensional curves.

\begin{table}[tb]
    \centering
    \small
    \begin{tabular}{l c} 
    \hline
        Variable & Descriptive statistics \\
        \hline
        n. signals & 11,928 \\ 
        n. trajectories & 1,216 \\ 
        n. unique anonymous users & 745 \\ 
        n. signals per trajectory & [5 - 118] \scriptsize{mean=9.8 sd=9.33} \\
        distance per trajectory [km] \hspace{0.5cm} & [0.3 - 33.0] \scriptsize{mean=5.1 sd=3.42}\\
        n. signals per anonymous user & [5 - 232] \scriptsize{mean=16.0 sd=23.12}\\
        n. trajectories per anonymous user & [1 - 11] \scriptsize{mean=1.6 sd=1.35}\\
        \hline
    \end{tabular}
    \caption{Descriptive statistics of the GPS dataset under analysis} 
    \label{tab:GPS_stats}
\end{table}

\subsection{Applying parameter optimization} \label{subsec:tuning_res}

\begin{figure}[tbh]
    \centering
    \includegraphics[width=8cm]{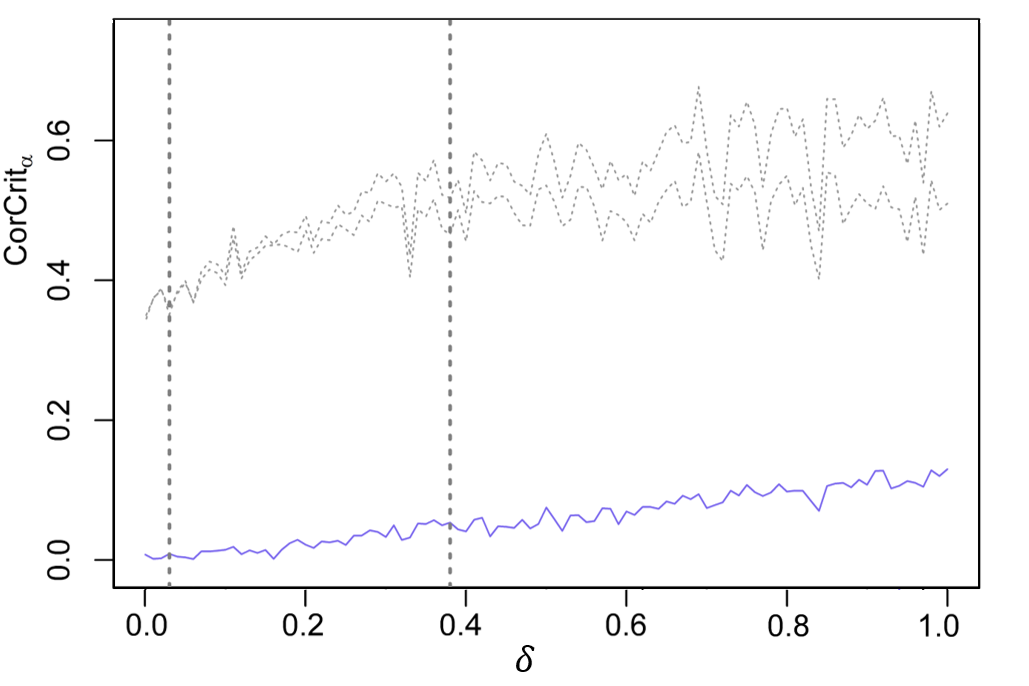}
    \caption{Plot of the cophenetic correlations (in grey) as $\delta$ varies in $[0,1]$. Their absolute difference is depicted in purple.}
    \label{fig:tuning_delta}
\end{figure}

The algorithm is tuned to identify the best values for the input parameters for the study case. Regarding the $\delta$ parameter, an in-depth analysis of the variation in the cophenetic index is conducted. Results indicate that there is a redundancy of information since the amplitude and phase distances give dendrograms that are already very close. In fact, as shown in Figure \ref{fig:tuning_delta}, the correlations between the combined distance and the two other distances (grey dotted curve) are close for all values of $\delta\in[0,1]$ and so the absolute difference (purple curve) is small and relatively constant. Any $\delta\in[0,1]$ is optimal to maximize the compromise of information from the amplitude and phase distances. $\delta=1$ is choosen among all possible values for the easy interpretation of its application. Setting $\delta=1$ in Equation \ref{eq:Dtot}  simplifies the total distance to the amplitude distance alone, effectively neglecting the phase component. However, this does not result in a loss of phase information, as the phase distance is so closely correlated with the amplitude that it is effectively still captured.

\begin{figure}
    \centering
    \subfigure[Clustering dendrogram]{\includegraphics[width=6cm]{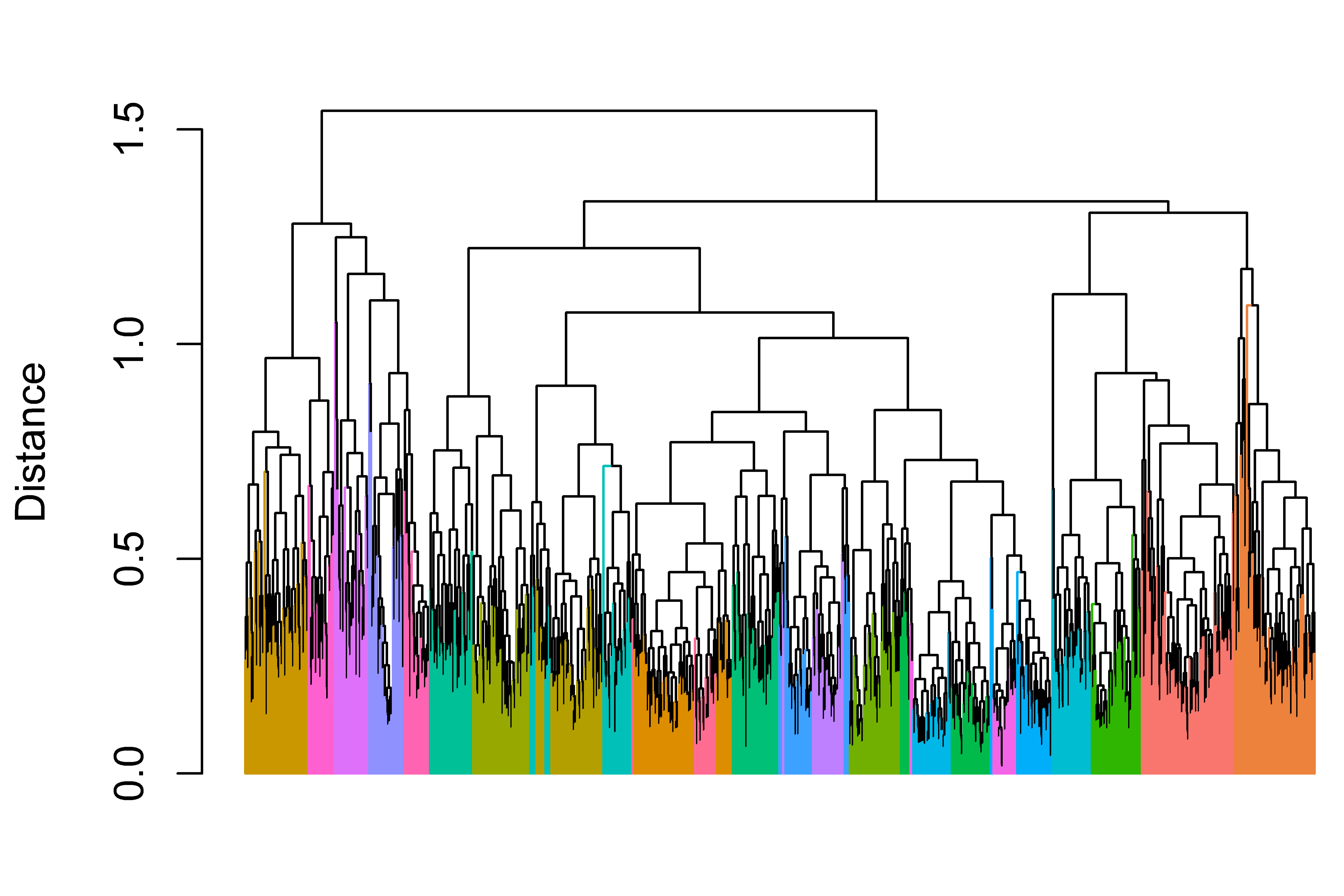}} 
    \hspace{0.5cm}
    \subfigure[Cluster sizes]{\includegraphics[width=6cm]{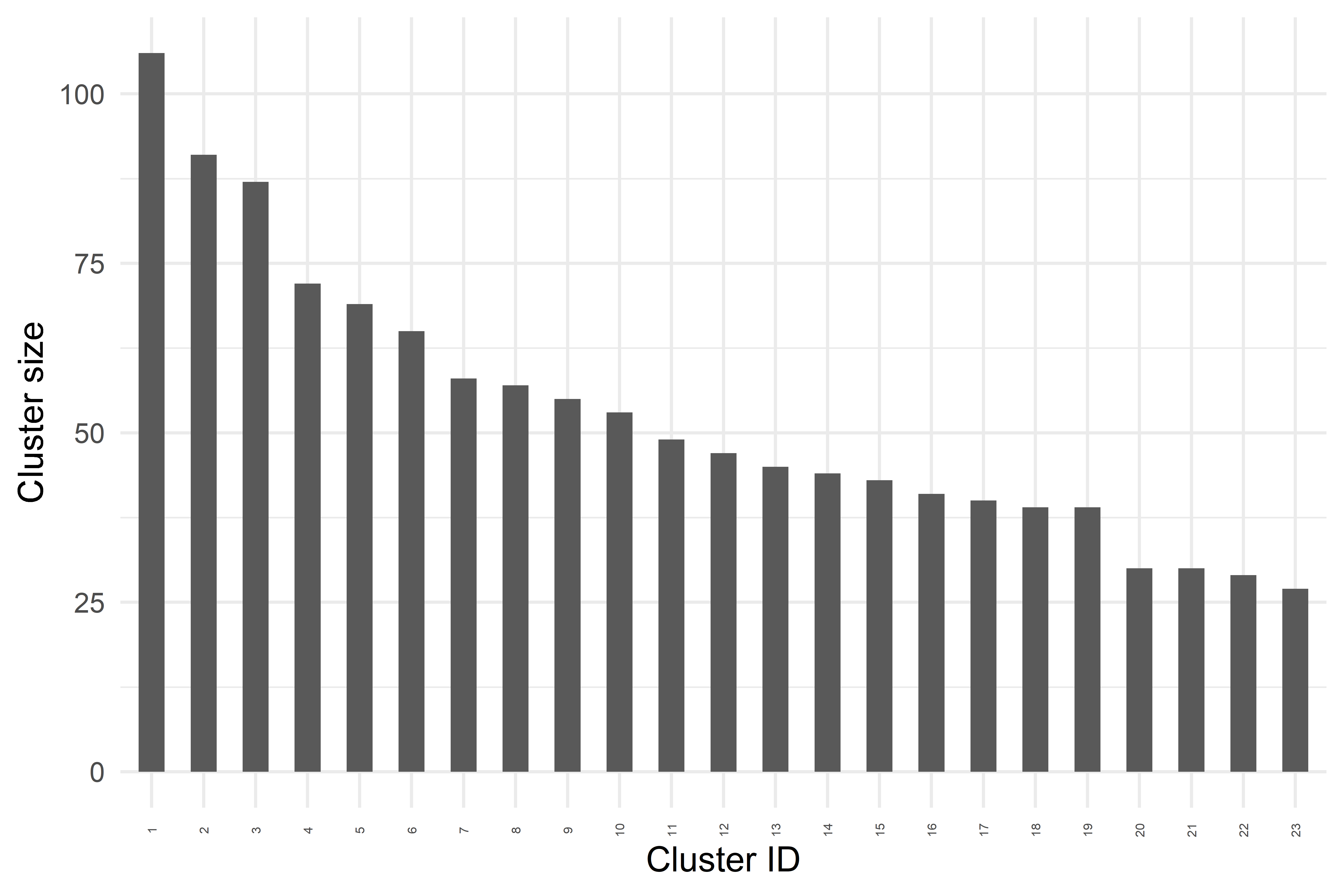}}
    \caption{On the left, dendrogram of the hierarchical clustering, colored by the output clusters. On the right, barplot of the cluster dimensionality.}
    \label{fig:clustering}
\end{figure}

The other two parameters, $K$ and $\alpha_0$, are set following the privacy and utility balance technique described in Algorithms \ref{scheme:tuning1} and \ref{scheme:tuning2}. At first, a total of $G=23$ groups of curves are identified using hierarchical clustering with dynamic cut. Their sizes range from $106$ to $27$, as shown in Figure \ref{fig:clustering}.

\begin{figure}
\centering
    \subfigure[Curves, grouped by $K$, represents the variation of the minimum distance index $\bm{I}^{(1)}$ over $\alpha_0$. The horizontal dotted line reports a possible fixed distance threshold.]{\includegraphics[width=6.5cm]{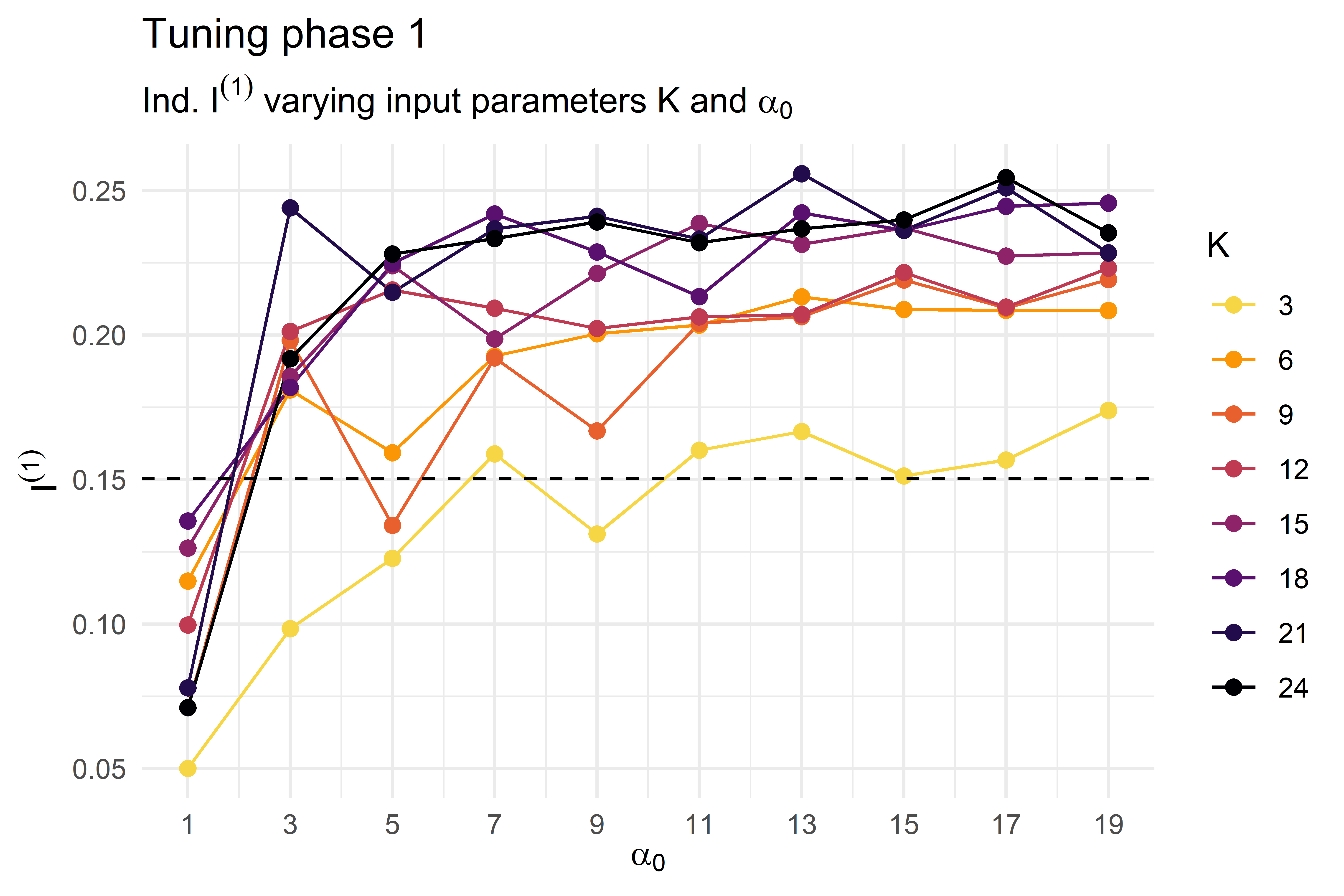}}
    \hspace{0.5cm}
    \subfigure[Boxplot of the absolute correlations of distances within groups for various $K$s. The straight black line depicts the variation of the indicator $\bm{I}^{(2)}$ with $K$.]{\includegraphics[width=6.5cm]{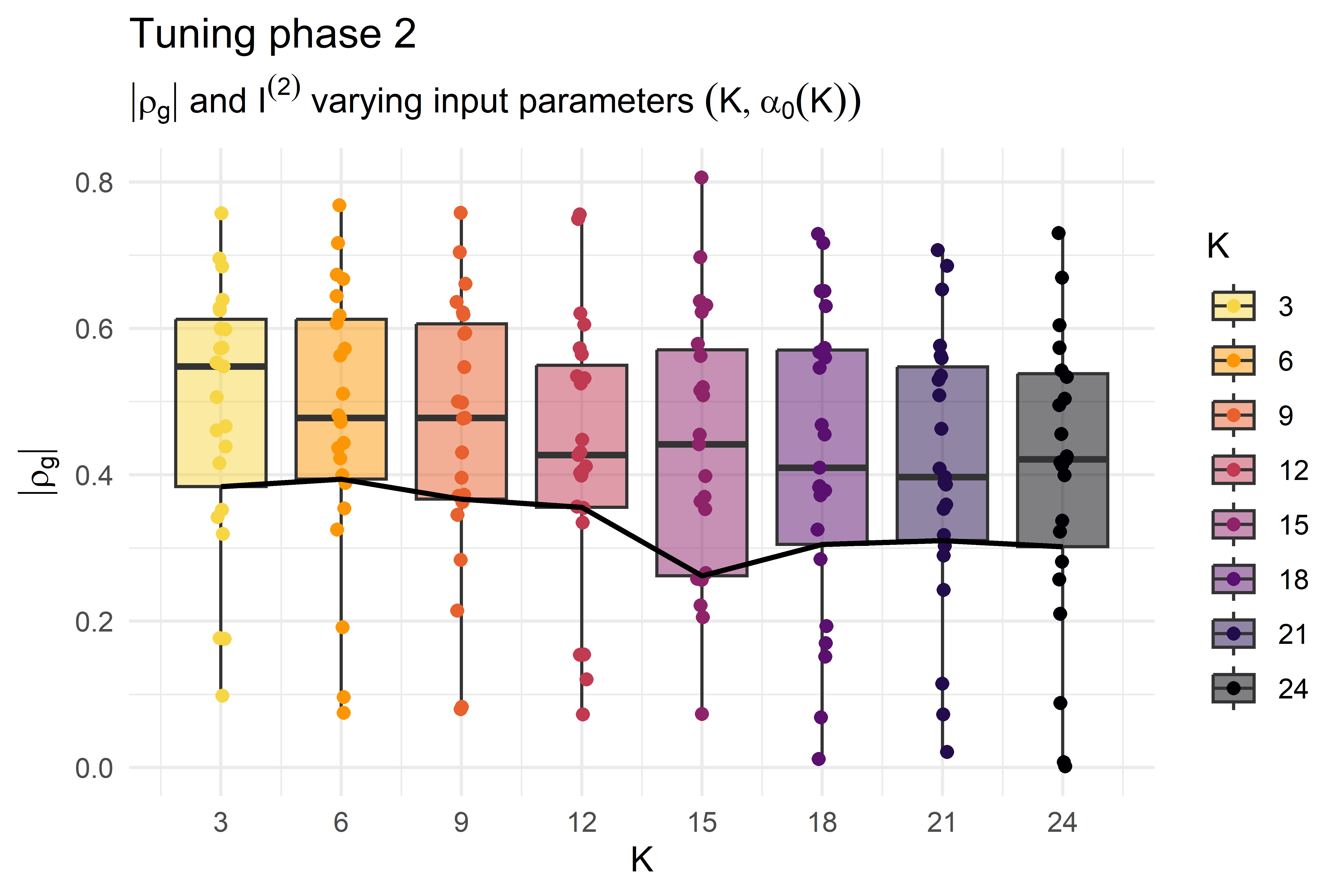}}
    \caption{Output of the first (left) and second (right) phases of tuning procedures for the choice of, respectively, $\hat{\alpha}_0(K)$ and $\hat{K}$.} 
    \label{fig:tuning_alpha_k:1}
\end{figure}

Eight values of $K$, equally spaced between 3 and 24, are tested, representing percentages from 0.25\% to 2\% of the number of the trajectories. For $\alpha_0$, ten values greater than 1 are tested, ranging from 1 to 19. For all combinations of $K\in\{3,6,\dots,24\}$ and $\alpha_0\in\{1,3,\dots,19\}$, the indicators $\bm{I}^{(1)}$ of Equation \ref{eq:tuning:i1} are computed and here reported in Figure \ref{fig:tuning_alpha_k:1}(a). Each of these curves illustrates the variation of the indicator $\bm{I}^{(1)}$ as $\alpha_0$ changes for each fixed value of $K$.
Each curve exhibits a rapid growth up to a certain point, identifiable as an ``elbow" point, after which the increase slows down and the curve stabilizes. The optimal value of $\alpha_0$ is determined for each value of $K$ as the point at which the stabilization occurs and is reported in the Table \ref{tab:tuning:1}. 
\begin{table}[tbh]
    \centering
    \begin{tabular}{|l|cccccccc|}
    \hline
        $K$ & 3 & 6 & 9 & 12 & 15 & 18 & 21 & 24 \\
        \hline
        $\hat{\alpha}_0(K)$ & 7 & 7 & 11 & 3 & 5 & 7 & 3 & 5 \\
    \hline
    \end{tabular}
    \caption{Optimal values of $\alpha_0$ for each value of $K$. The results are reported when using the elbow method, i.e., by chosing $\alpha_0$ as the point of curve stabilization.}
    \label{tab:tuning:1}
\end{table}
The analysis of the variability of the correlations $\{|\rho_g|\}_g$ as $(K, \hat{\alpha}_0(K))$ changes is shown in Figure \ref{fig:tuning_alpha_k:1}(b). In these plots, each dot represents the the single absolute correlations $|\rho_g|$ within groups, while the straight lines indicate how the indicator $\bm{I}^{(2)}$ varies with $K$. As observed, the maximum value of the indicator is reached for $K=6$. The final optimal parameters used in the synthesis procedure will hence be $K=6$ and $\alpha_0=7$.

\subsection{Qualitative evaluations}\label{subsec:qualitative}

\begin{figure}
    \centering
    \subfigure[Original (sample A)]{\includegraphics[width=4cm]{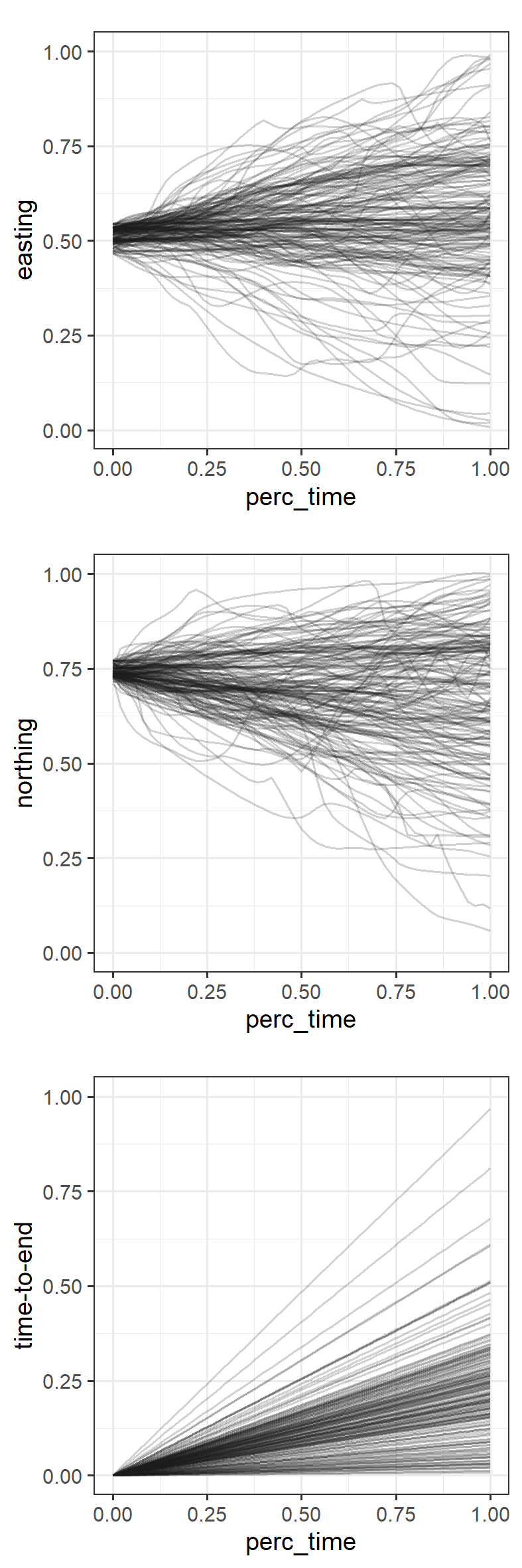}}
    \hspace{0.25cm}
    \subfigure[Synthetic (sample A)]{\includegraphics[width=4cm]{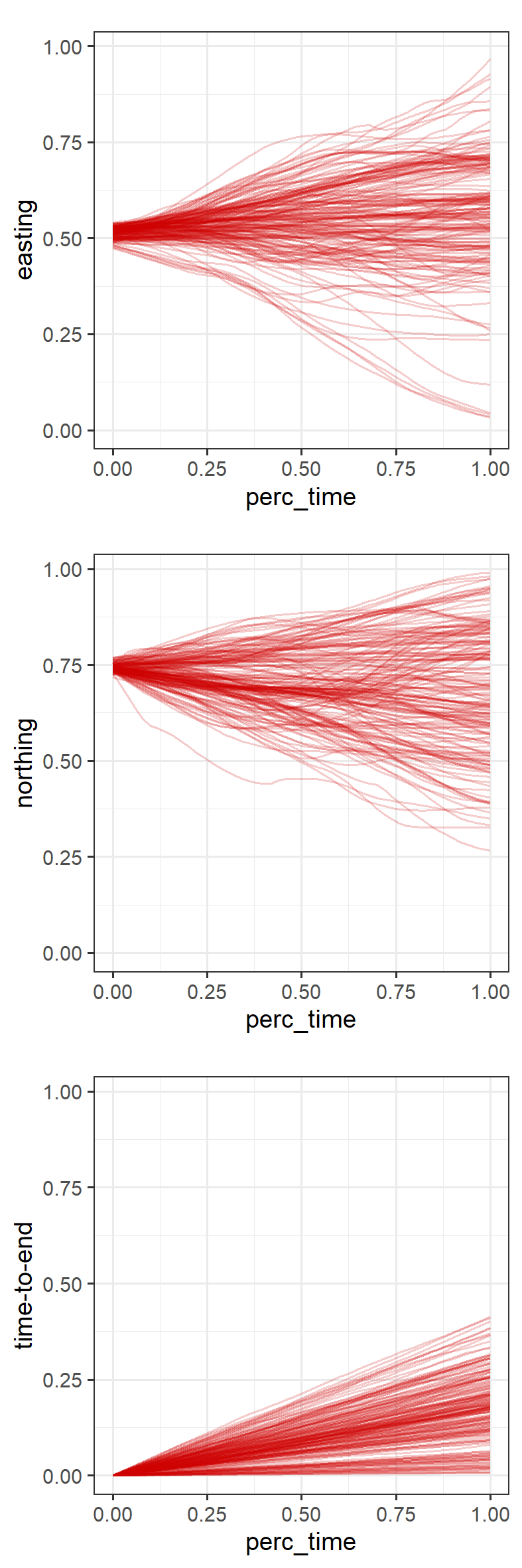}}
    \hspace{0.25cm}
    \subfigure[Synthetic (sample B)]{\includegraphics[width=4cm]{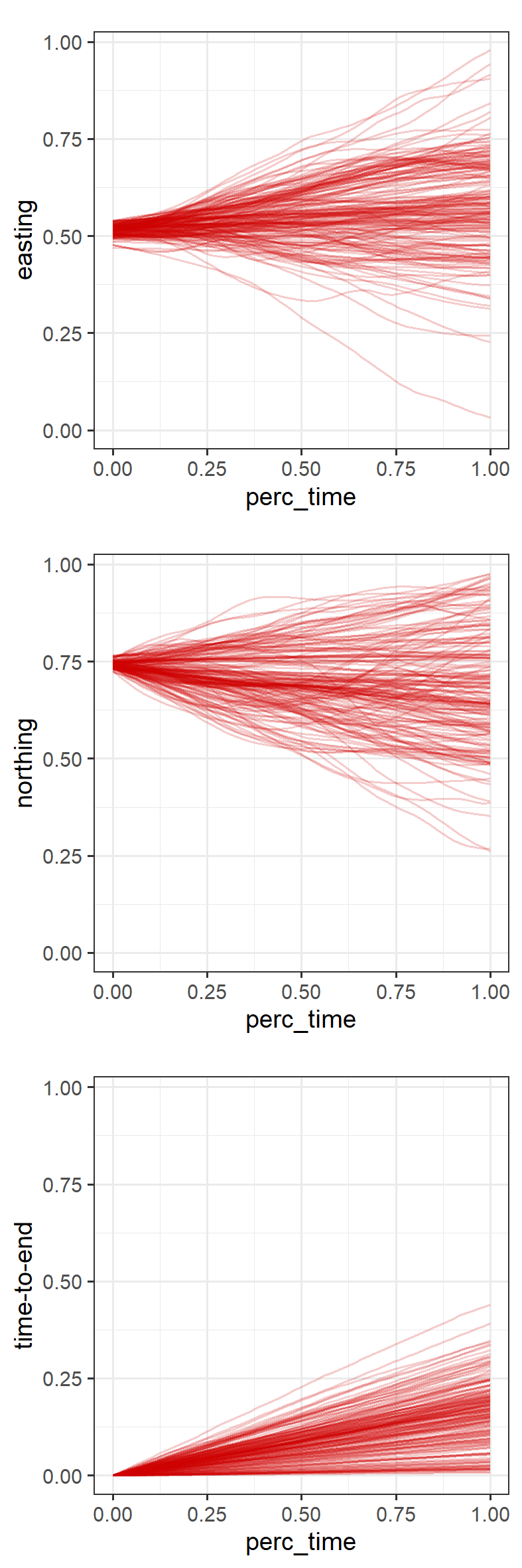}}
    \caption{Representation of some functions, both original (left) and synthetic (center and right).}
    \label{fig:viz:functional}
\end{figure}

To assess the effectiveness of the proposed methodology, the performance is initially evaluated from a qualitative approach. The qualitative analysis allows for a visual examination of the similarity between synthetic and original trajectories, ensuring that the generated data preserves both global and local characteristics of the mobility patterns.
First, the synthetic trajectories are compared with the original ones, treating them as multidimensional functions, in line with the approach used to model the input and output in the FDASynthesis method. Figure \ref{fig:viz:functional} illustrates the representation of the multivariate functions, divided into three sections: the original functions (left section), the synthetic functions corresponding to the same original curves (middle section), and an additional random sample of synthetic functions (right section). To avoid confusion due to the large size of the dataset, a random subsample of 15\% of the total trajectories is selected. In the middle section, the synthetic curves are associated with the same subsample of original curves, ensuring a direct one-to-one comparison between original and synthetic data. In the right section, a different random sample of 15\% of the trajectories is used. By examining this second sample, any patterns of consistency or variation between different synthetic subsets can be observed, offering further insight into the robustness of the method across different data. In each section, the three components of the functions are represented in the domain $[0,1]$: easting (top), northing (middle), and elapsed time (bottom). 
From this visualization, it is evident that the local properties of the original functions are preserved. Well-defined mobility channels are clearly observable and these are maintained in the synthetic dataset as well. However, it appears that the variability of the original data is somewhat reduced: the synthetic curves look smoother, and extreme values such as higher time durations tend to be diminished, indicating the absence of the distribution tails.

\begin{figure}
    \centering
    \subfigure[Original (sample A)]{\includegraphics[width=4cm]{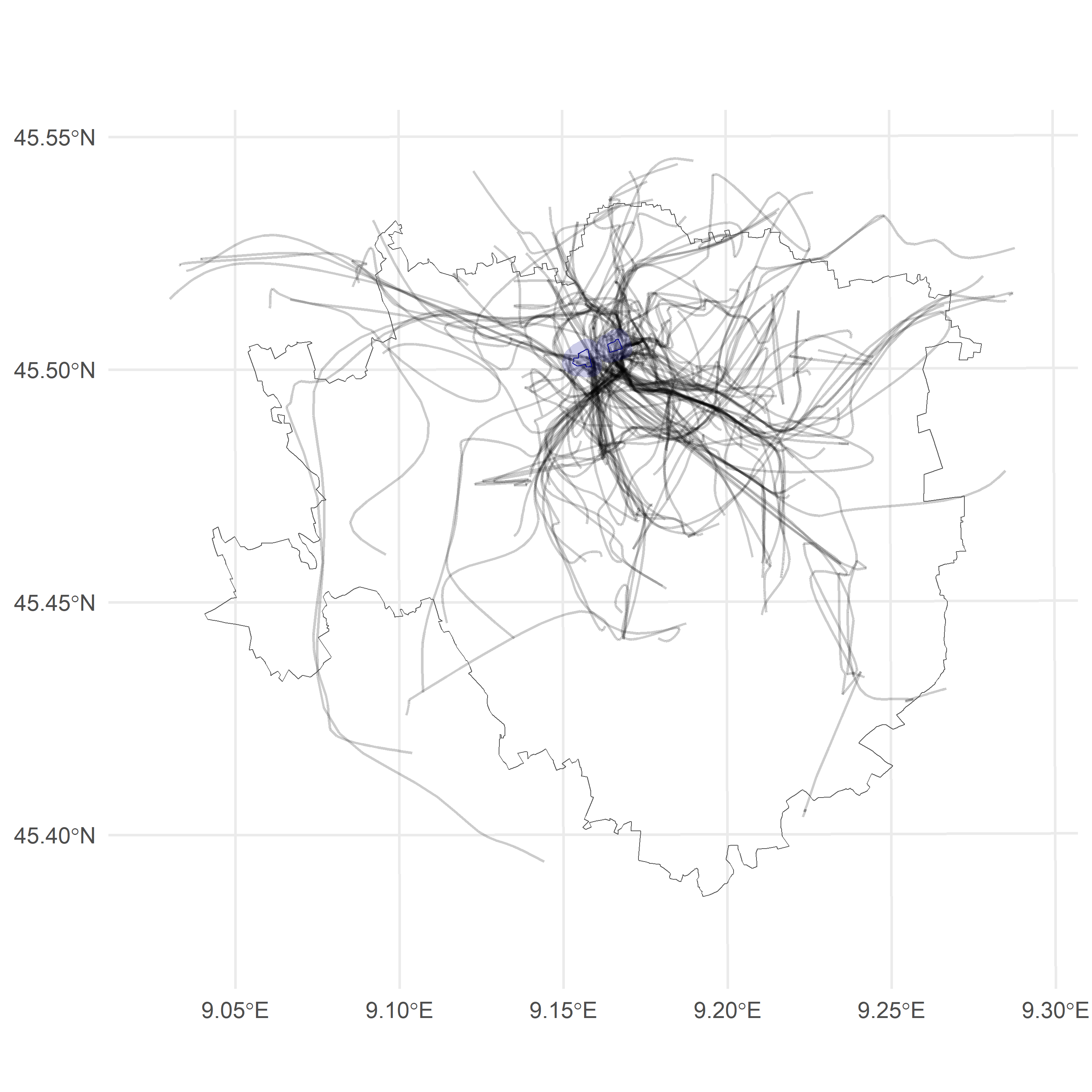}}
    \hspace{0.5cm}
    \subfigure[Synthetic (sample A)]{\includegraphics[width=4cm]{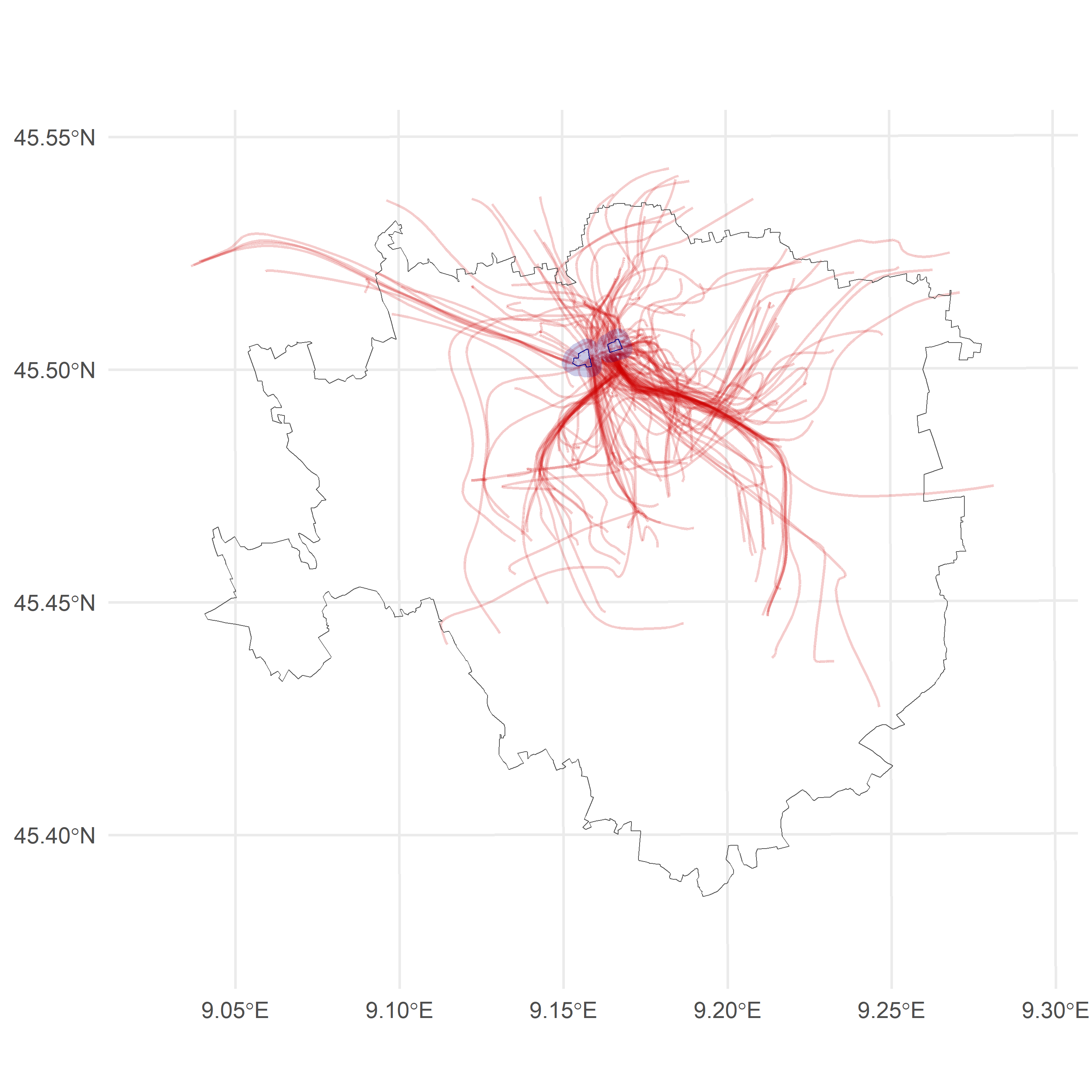}}
    \hspace{0.5cm}
    \subfigure[Synthetic (sample B)]{\includegraphics[width=4cm]{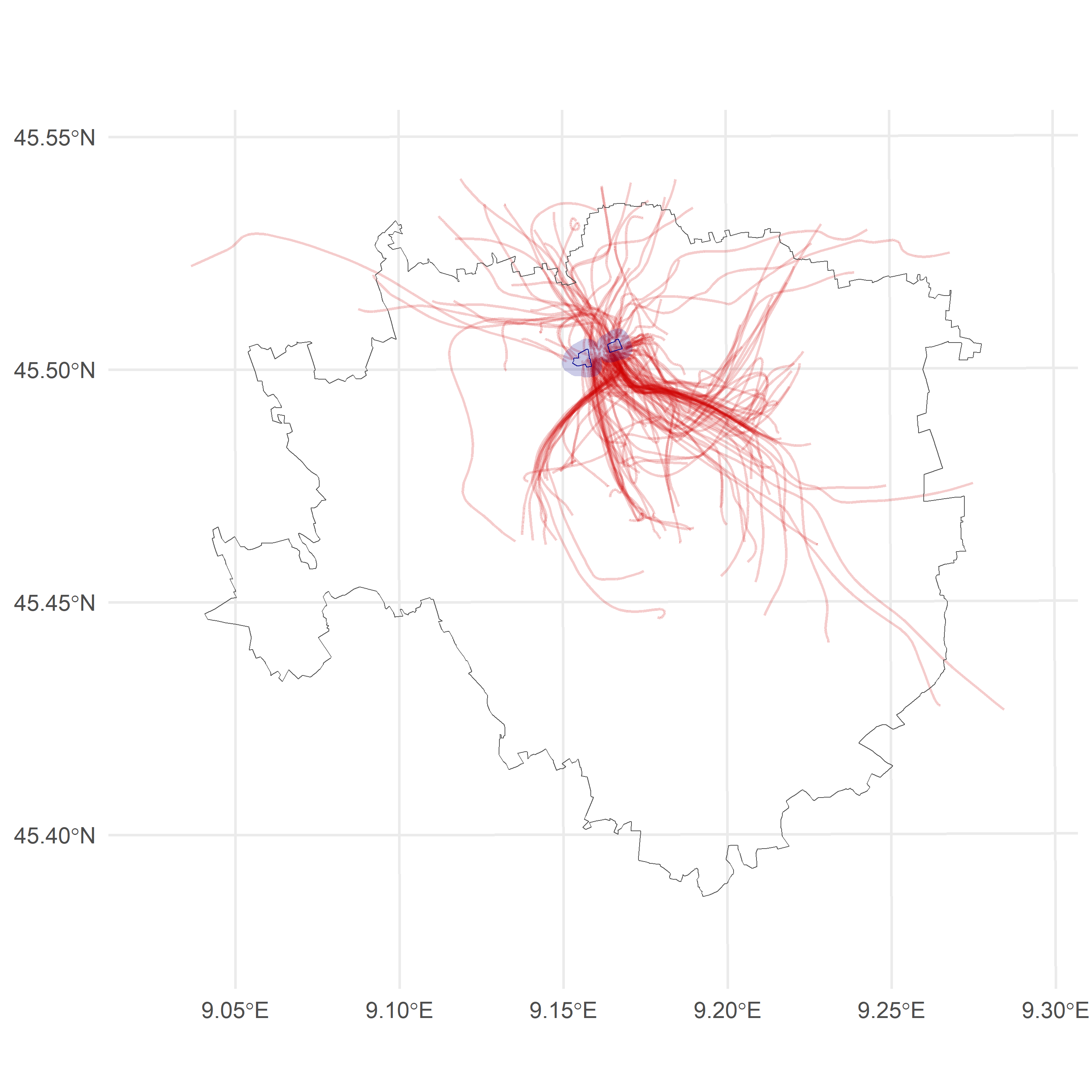}}
    \caption{Projection of some functions, both original (left) and synthetic (center and right), on the original map of Milan. The Bovisa campus is highlighted in blue. The black border represents the boundary of the Milan area.}
    \label{fig:viz:map}
\end{figure}

Similar conclusions can be drawn from the second visualization, shown in Figure \ref{fig:viz:map}. In this case, the functions are projected onto the physical map, illustrating the spatial movements in the metropolitan area of Milan, although without the temporal component. As in the previous visualization, there are three sections representing a plot of a subsample of $15$\% of the original trajectories, the same subsample of synthetic trajectories, and a different but equally-sized subsample of synthetic trajectories. 
Once again, well-defined mobility channels are visible and preserved in the synthetic dataset. For example, the trajectories along the ring road and those on the roads coming from the north-west are clearly maintained. This indicates that the synthetic data successfully replicates the primary traffic flows and movement patterns that are characteristic of the original data.
However, not all areas of Milan are covered as extensively in the synthetic data as in the original. Some regions in the original dataset appear underrepresented, like the south-west and the south-north areas, presumably due to the smoothing of less frequent patterns or exclusion of rarer trajectories. While major traffic routes are well-preserved, areas with lower traffic density or sporadic movement may not be as accurately replicated.  

\begin{figure}
    \centering
    \subfigure[Original (sample A)]{\includegraphics[width=4cm]{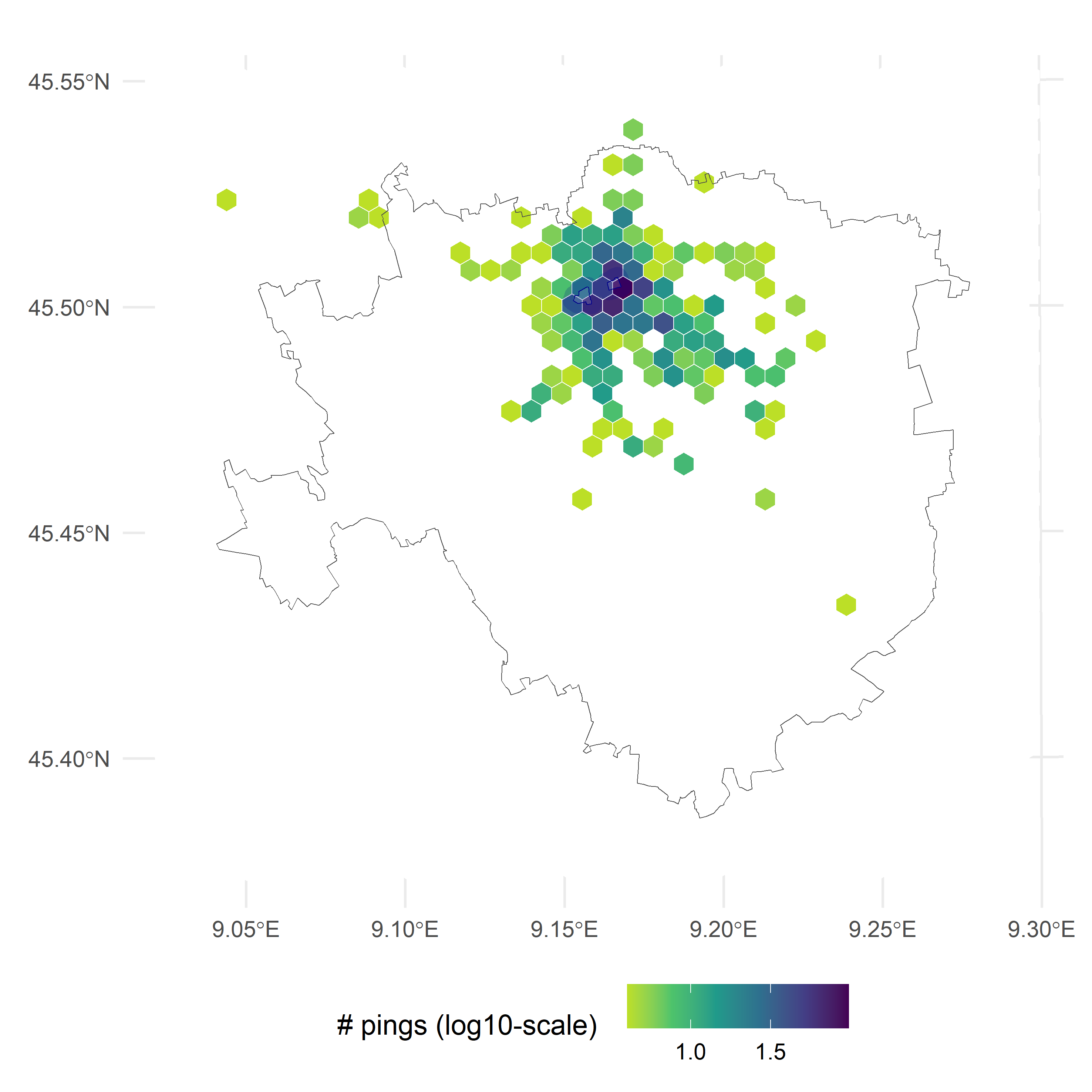}}
    \hspace{0.5cm}
    \subfigure[Synthetic (sample A)]{\includegraphics[width=4cm]{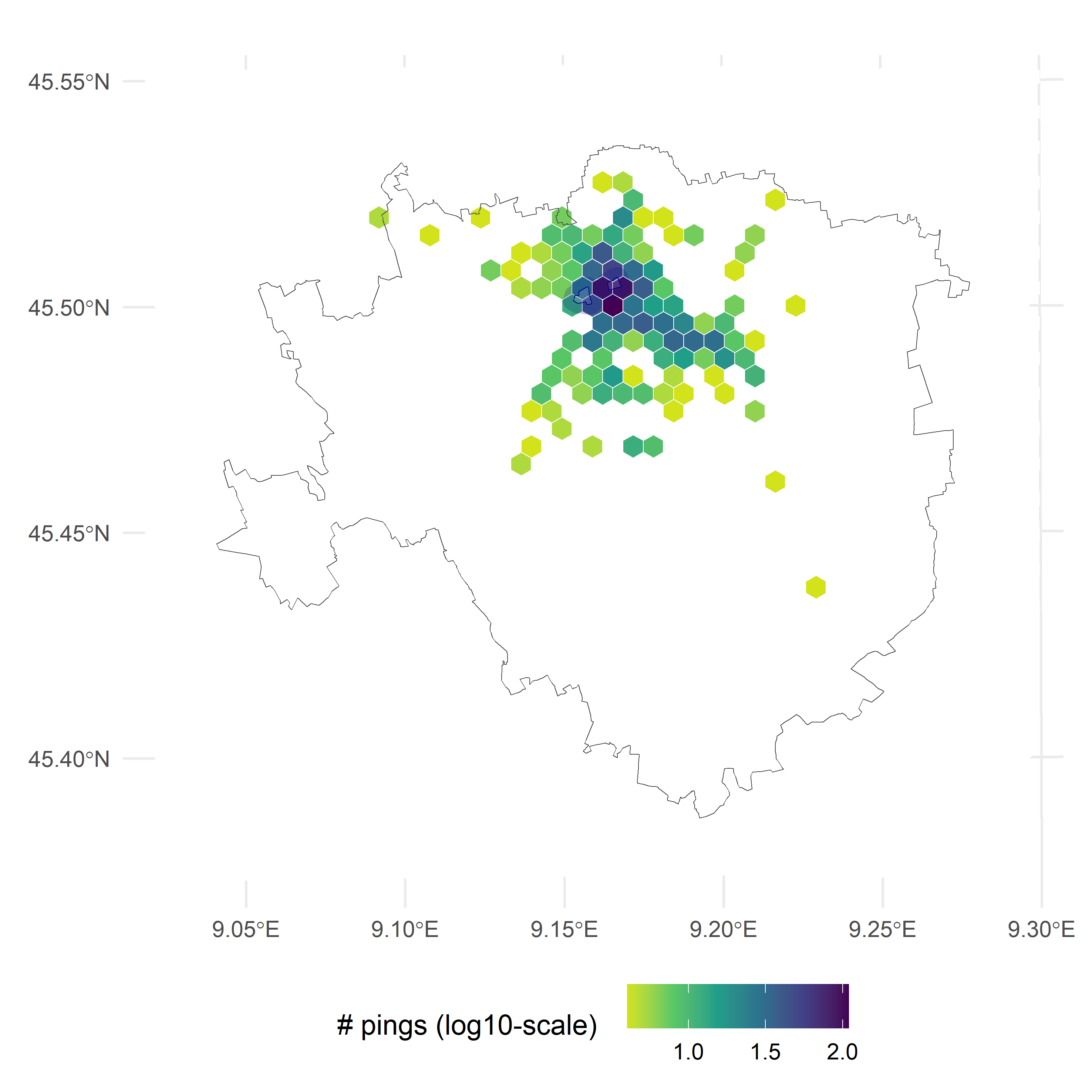}}
    \hspace{0.5cm}
    \subfigure[Synthetic (sample B)]{\includegraphics[width=4cm]{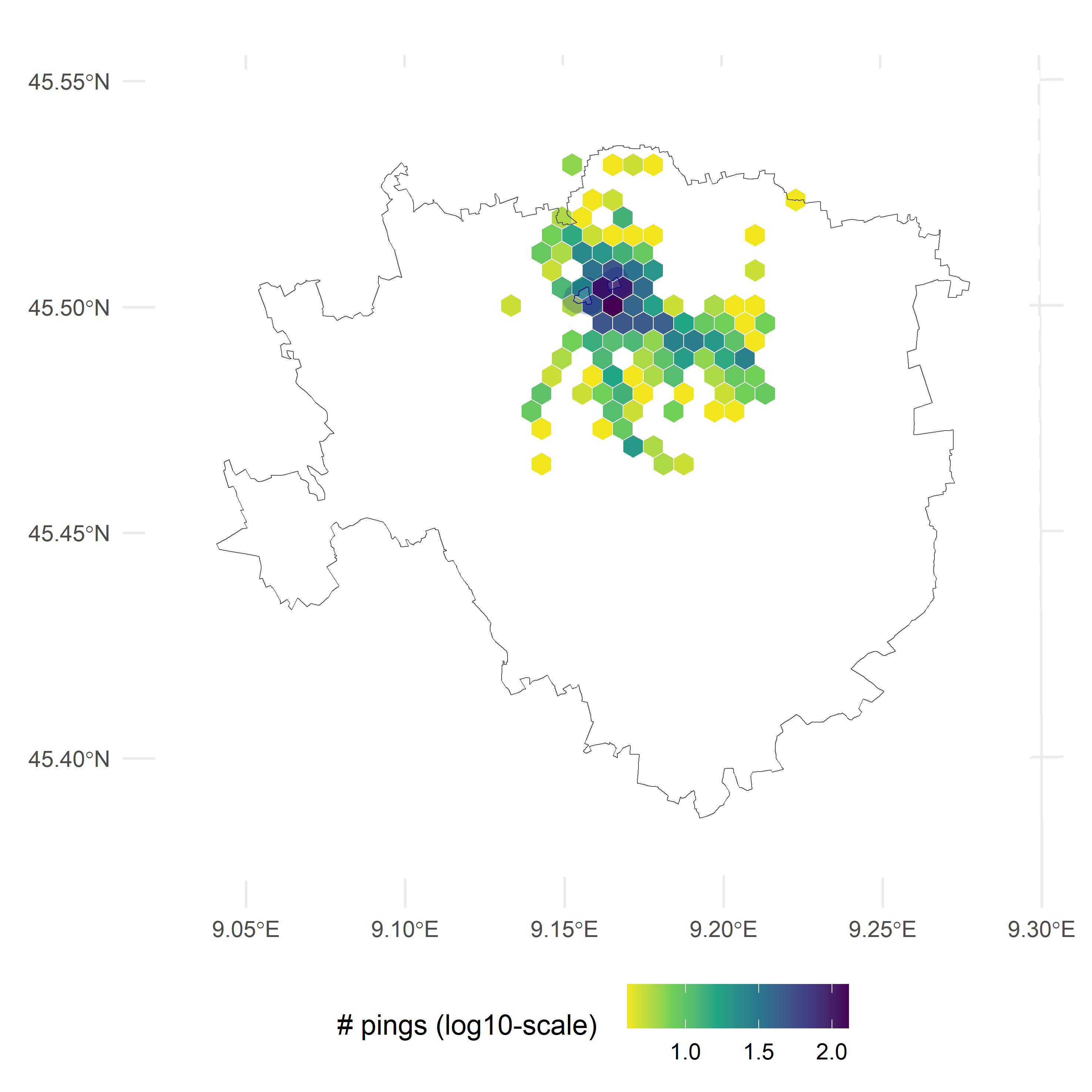}}
    \caption{Heatmaps of the most visited neighbors after subsampling.}
    \label{fig:viz:heatmap}
\end{figure}

The third and final plot analyzes the number of visits per area. Points are sampled from the functions, converting the functional model back into a discrete structure similar to raw GPS data. The area of interest is divided into equally sized hexagonal cells of diagonal $\sqrt{3}/3$ km, and the number of points in each zone is counted. This check is crucial to ensure that the aggregate number of visitors is comparable between original and synthetic data. Figure \ref{fig:viz:heatmap} shows the three heatmaps: one for the original subsample, one for the synthetic subsample, and one for another synthetic subsample. The results are consistent across all maps. Despite the variations observed in the peripheral areas, the overall distribution of visits remains similar. This observation indicates that the synthetic model faithfully replicates the core visit patterns found in the original data.

\subsection{Quantitative evaluations}\label{subsec:quantitative}

\begin{figure}
    \centering
    \subfigure[Test for the mean functions]{\includegraphics[width=6.5cm]{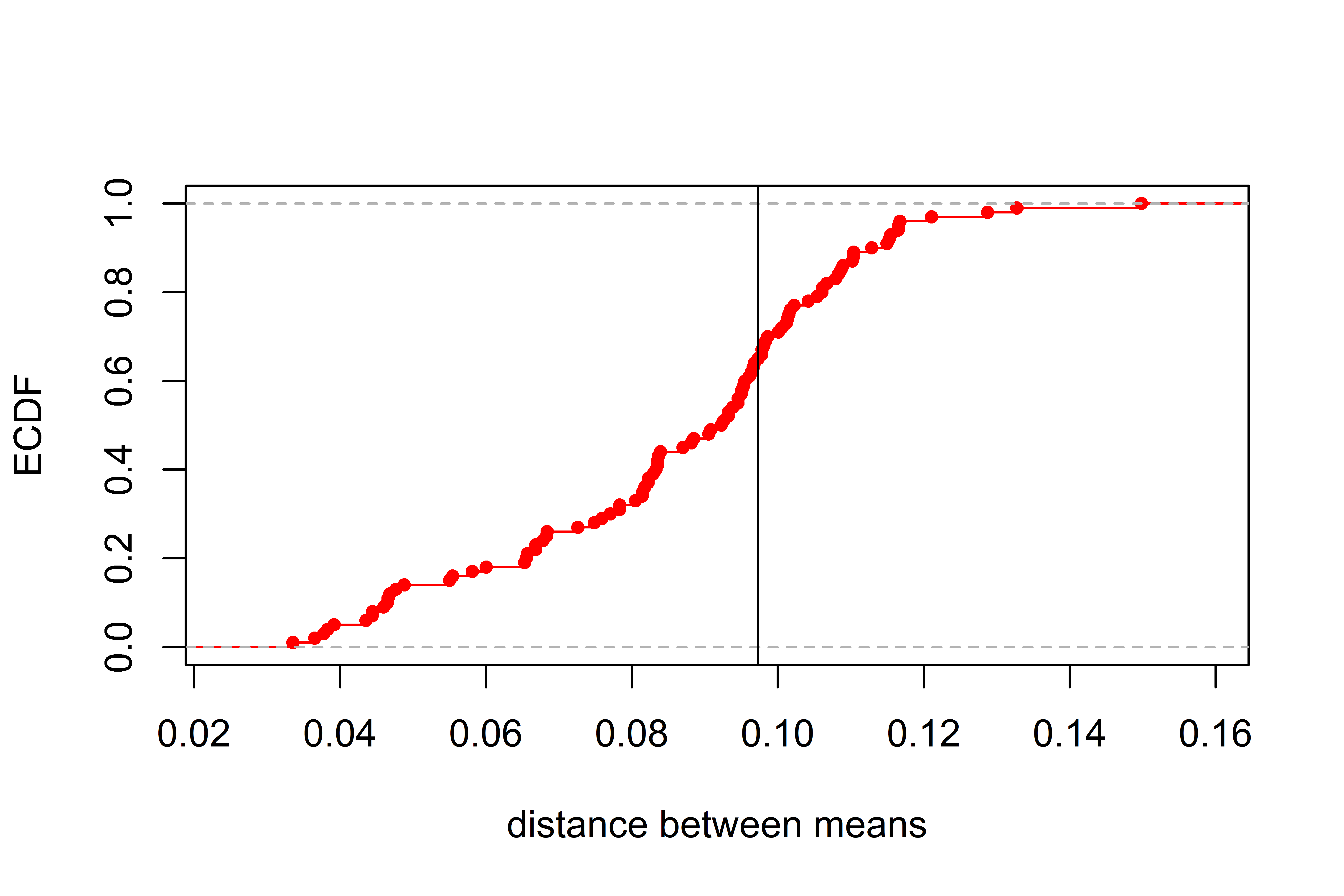}}
    \hspace{0.5cm}
    \subfigure[Test for the covariance operators]{\includegraphics[width=6.5cm]{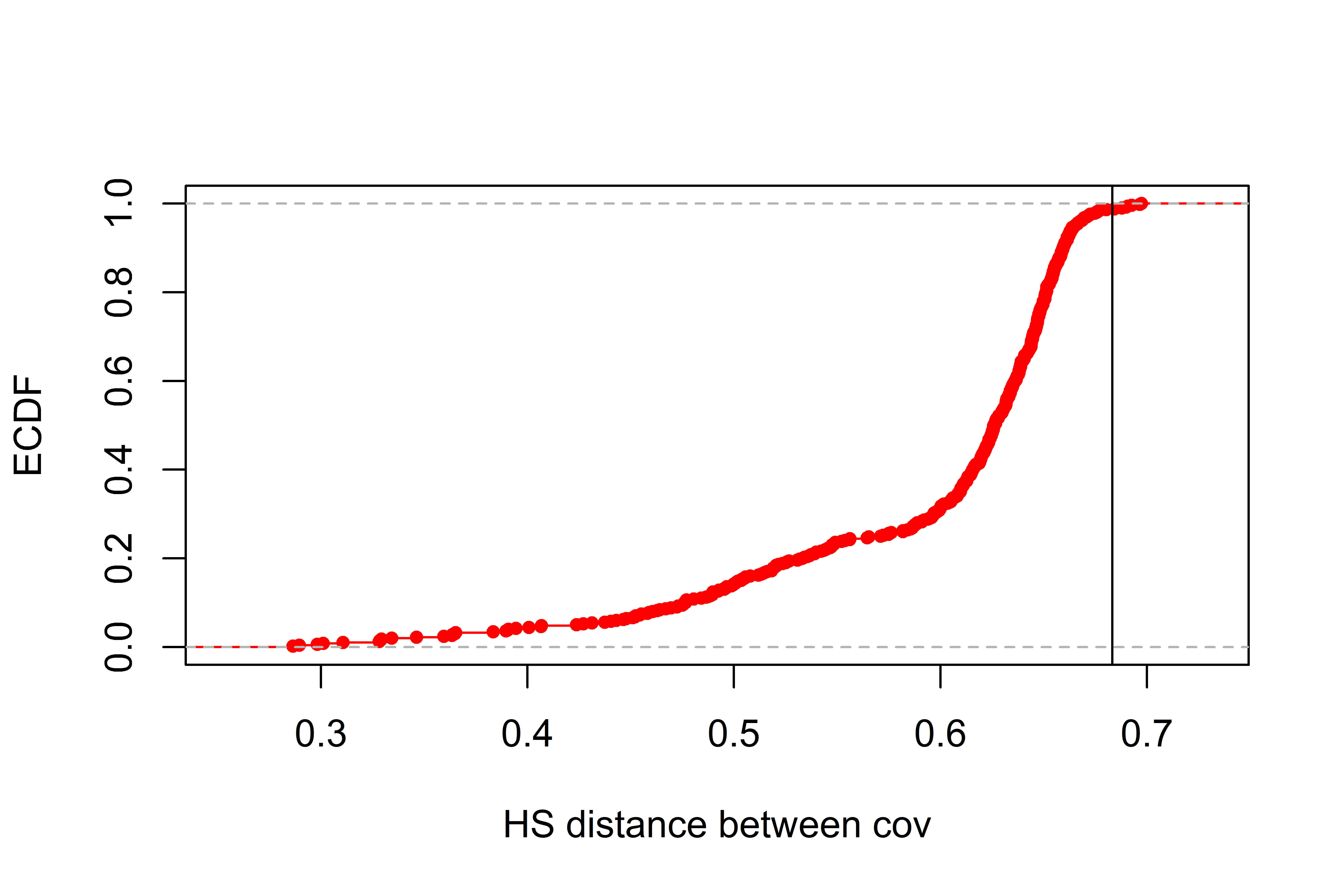}}
    \caption{On the left, ECDF of the distances between mean functions. On the right, ECDF of the distances between covariance operators. In both plots the vertical line represents the observed test statistic.}
    \label{fig:test1}
\end{figure}

After completion of the visual assessment phase, it is essential to proceed with a quantitative analysis of the resemblance between the original and synthetic datasets. This process requires a rigorous evaluation to ensure that the synthetic dataset not only visually resembles the original dataset but also maintains fundamental statistical properties. Two specific statistical tests have been applied to compare and confirm the equality of the moments of the distributions of the two datasets. 
The first test focuses on the equality of the means of the distributions, namely the functions $\bm{\mu}_f\in\mathcal{F}(I)$ for the original dataset and $\bm{\mu}_f^s\in\mathcal{F}(I)$ for the syntehtic one, defined as in Equation \ref{eq:meanf}. The means are compared through a permutational test in which the test statistics are the distance between the two mean functions of the sample, as defined in Equation \ref{eq:Dtot}, and it is tested whether this distance is zero or greater. The following Equation \ref{eq:test1} reports the null and alternative hypotheses of the test.
\begin{equation} \label{eq:test1}
\begin{split}
    H_0: \quad & D(\bm{\mu}_f, \bm{\mu}_f^s) = 0 \\
    H_1: \quad & D(\bm{\mu}_f, \bm{\mu}_f^s) > 0 \\
\end{split}
\end{equation}
The test is performed with $500$ permutations, shuffling the data across the two samples, and generates a distribution of the test statistic whose Estimated Cumulative Distribution Function (ECDF) is reported in Figure \ref{fig:test1}(a). The empirical p-value $p_{\mu}=0.36$, calculated as the proportion of permutations that yielded a test statistic greater than the observed value, is large enough to support the equality of the means.

The second test deals with the equality of covariance operators. The aim is to measure the significance of the difference between the two covariance operators, $\bm{\Sigma}$ for the original dataset and $\bm{\Sigma}^s$ for the synthetic dataset, using the concept of distance between covariance operators $d(\cdot, \cdot)$. Following the work by \citet{Pigoli2014Distances}, various distance metrics are explored in this research, namely the Hilbert-Schmidt norm, the operator norm, the trace norm, and the square-root transformation. Since they have always led to consistent results, the results presented here refer only to the Hilbert-Schmidt metric. The hypotheses of the statistical test are formulated as shown in Equation \ref{eq:test2}.
\begin{equation}\label{eq:test2}
\begin{split}
    H_0: \quad & d(\bm{\Sigma}_f, \bm{\Sigma}_{f}^s) = 0 \\
    H_1: \quad & d(\bm{\Sigma}_f, \bm{\Sigma}_{f}^s) > 0 \\
\end{split}
\end{equation}
A permutation test is used, employing the distance between covariance operators as the test statistic and $500$ random permutations of the labels on the two sets of curves. Since the means of the two datasets are unknown and different from zero, at each iteration, before calculating the sample covariance operator, the curves are centered using their sample mean. The ECDF of the test statistic, shown in Figure \ref{fig:test1}(b), and the p-value of the test, $p_{\Sigma}<0.001$, suggest rejecting the null hypothesis. Therefore, the covariance operators of the two original and synthetic datasets are statistically different.

\begin{figure}
    \centering
    \includegraphics[width=6cm]{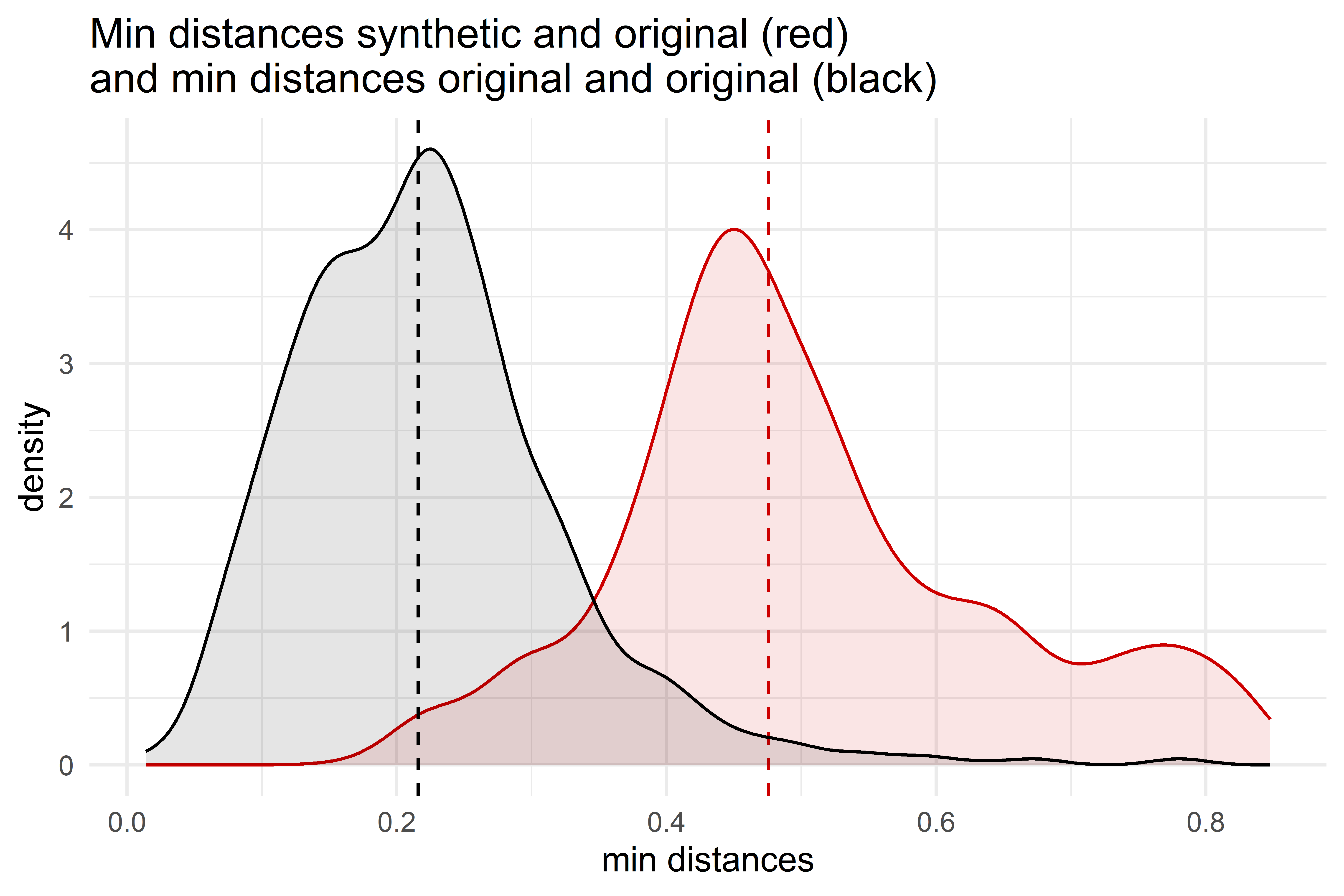}
    \caption{Distribution of the minimum distances among original functions (in black) and between original and synthetic curves (in red). The vertical dotted lines represent the value of the median of the distributions.}
    \label{fig:privacy_check:1}
\end{figure}

In addition to evaluating the similarity between the original and synthetic datasets using both quantitative and qualitative measures, it is essential to ensure that the synthetic data generation process adheres to privacy standards. Ensuring that the method protects individual privacy while maintaining data utility is a key aspect of its effectiveness.
The choice of the parameters $K$ and $\alpha_0$ already takes into account the respect of the privacy within the FDASynthesis process. A final check is implemented, looking at the minimum distances between original and synthetic functions, to ensure that the 1-to-1 correspondance between synthetic and original data is avoided. Figure \ref{fig:privacy_check:1} shows in red the distribution of the minimum distances between the original curves and the synthetic ones. The black distribution is, instead, related to the distances among the original functions themselves. The vertical lines represent the median values of the synthetic-original distance distribution and the original-original one, equal to $0.476$ and $0.216$, respectively.
This graph confirms that privacy is respected since the minimum distances between synthetic and original data are on average twice larger than the minimum distances between the original and themselves. Indeed, it is more likely for a real trajectory to carry sensitive information and risk to expose another original datum compared to a synthetic trajectory.

\subsection{Comparative analysis of a different parameter setting}

All the results presented in Subsections \ref{subsec:qualitative} and \ref{subsec:quantitative} refer to the process that uses the input parameters $\delta=1$, $K=6$, and $\alpha_0=7$, chosen as described in Subsection \ref{subsec:tuning_res}. However, the choice of the last two parameters $K$ and $\alpha$ can be adjusted to generate a dataset that is less focused on privacy compliance and more oriented toward utility, using a method other than the data-driven elbow search. If a more relaxed or stricter privacy restriction is desired, this can be achieved by setting a minimum threshold for the indicator $\bm{I}^{(2)}$ and selecting, for each value of $K$, the first $\alpha_0$ values that exceed this threshold as the optimal values. For example, a lower threshold that relaxes privacy requirements could be a distance threshold of $0.150$, corresponding to the $25$-th percentile of the distribution of minimum distances between the original curves. Different values of $\hat{\alpha}_0(K)$ and consequently different possible values of $\hat{K}$ are obtained, indeed $K=3$ and $\alpha_0=7$. The full results of this analysis are presented in Appendix \ref{sec:app1}, while here only the key findings are discussed in comparison with the previously analyzed synthetic data.

The synthetic data generated with this second pair of input parameters successfully replicates the structure and spatiotemporal dynamics of the original functional data, with trajectories closely matching real mobility patterns. Furthermore, the method preserves the aggregate mobility dynamics, evidenced by a comparable spatial distribution of visit frequencies. However, compared to the previously analyzed case, this time, a higher variability in the data is maintained. The choice of a lower $K$ leads to results more closely aligned with specific observations and less influenced by smoothing effects, as was the case with the larger $K=6$. 
The preservation of the statistical characteristics of the datasets is assessed by comparing the mean functions and the covariance operators. This time, unlike before, both tests output the equality of the moments of the two datasets, resulting in a p-value of $0.16$ for the test on the means and $0.39$ for the test on the covariance operators. Hence, as suggested by the visual inspection, a better result in terms of similarity of the datasets is achieved with this second choice of parameters. 
Moreover, this improvement still ensures privacy at a level of security similar to before. This is evident when comparing the minimum distances between the original curves and the synthetic ones, as well as between the original curves themselves. The synthetic-original distance is still, on average, twice as large as the original-original one. Although this distance has now decreased, it is still high enough to provide a reasonable level of privacy.

\section{Discussion and conclusion}\label{sec:conclusions}

FDASynthesis is a synthetic data generation method for functional data that, taking as input a dataset of curves, synthesizes a dataset of the same number of functions. Specifically, the functions in the original dataset are described as elements of the space of absolutely continuous functions, and their respective SRVFs are calculated. By finding the optimal alignment between them, the amplitude and phase distances are evaluated. Such distances are then incorporated into a single distance measure, a convex combination of the two, in which the combination parameter $\delta$ is chosen after analyzing the concordance of the two sources of information. Having found the neighborhood of each original reference function, the associated synthetic function is computed as a weighted average, with stochastic weights, of the functions in the neighborhood. The size of the neighborhood, $K$, and the concentration parameter of the weights distribution, $\alpha_0$, are chosen after a parameter tuning step aimed at finding the right balance between maintaining the similarity of the synthetic dataset to the original and ensuring compliance with privacy constraints. 

Although FDASynthesis recalls the $k$-nn method, as both construct a weighted average around each original curve, the former goes beyond determinism by introducing a stochastic element into the calculation of the averages. This stochastic aspect is fundamental because it allows for the introduction of variability and controlled noise into the calculated averages, which results in enhanced privacy protection of the generated synthetic data.
Another key aspect is the handling of functions within a suitably chosen mathematical space, which allows for the definition of distances between functions in a more accurate and meaningful way. This is a crucial point, as the choice of mathematical space and the definition of the distance between functions directly influence the quality and utility of the generated synthetic data.
These innovations make FDASynthesis a cutting-edge approach in the field of functional data synthesis, with potentially broad applications in sectors requiring the generation of secure and representative synthetic data.

In the current research, the method has been applied to a dataset of GPS signals collecting trajectory movements of anonymous users from a common point of interest to any point the Milan metropolitan area. GPS trajectories are modeled as functions with three dimensions. After proper pre-processing and smoothing, synthetic trajectory functions are obtained through FDASynthesis starting from the original ones. The analysis of the results reveal that such synthetic data are able to capture the main features of the original functions and, when projected on the physical map, highlight similar mobility patterns. Statistical tests on the equality of the mean and covariance structure reveal that FDASynthesis is able to preserve such structures up to a suitable choice of the parameters. Additionally, the method ensures user privacy, as evidenced by privacy checks on minimum distances. While maintaining the relevant statistical characteristics of the original dataset, it introduces sufficient variability to ensure privacy.

As evident from the discussion of the results, the performance of FDASynthesis depends on the choice of hyperparameters, particularly $K$ and $\alpha_0$. Alongside FDASynthesis, this research also introduces a novel data-driven method for selecting parameters, designed to guide the parameter choice effectively. 
It has been observed that selecting a method with greater privacy focus (e.g., case of $K=6$, $\alpha_0=7$) yields sufficiently good results in terms of similarity, as the data retains many key structural patterns of mobility. However, there is also a reduction in coverage of more peripheral or less-traveled areas, caused by the intrinsic difference of the covariance operators of the two functional datasets. This phenomenon of reduced coverage is not necessarily a drawback, as it can enhance privacy. By excluding rare trajectories, the synthetic data reduces the risk of identifying individuals who, due to their unique and less common routes, might otherwise be more easily identifiable in the dataset.
However, if the main interest shifts to the utility of the synthetic data, relaxing the privacy constraints with a more permissive parameter choice (e.g., case of $K=3$, $\alpha_0=7$) allows FDASynthesis to achieve even greater similarity with the original data and to preserve the structure of the moments, at least up to the second moment.
This highlights the trade-off between the need to avoid the identifiablity of instances and to fully represent the diversity of functions of the original data. The balance between privacy and similarity is a critical consideration in the design and application of synthetic data generation methods like FDASynthesis. When the focus is on protecting privacy, certain unique or less common trajectories might be omitted or generalized to prevent re-identification, but this can lead to a reduction in the richness and representativeness of the synthetic data. In contrast, if the priority is to maintain a high degree of similarity to the original data, there is a greater risk of inadvertently preserving identifiable patterns. In applying the FDASynthesis method, it is essential to keep in mind the research goals to properly balance the two dimensions accordingly.

Further developments of this research could be directed in two main areas, each with the potential to deepen the understanding and expand the applicability of the FDASynthesis algorithm. On one hand, it would be valuable to test the algorithm on different types of functional data across various application domains, such as biometric signals, economic time series, or environmental data. Extending the analysis to diverse datasets could confirm the robustness and effectiveness of FDASynthesis, and potentially reveal optimizations needed for specific domains. Additionally, FDASynthesis is applicable to functional data characterized by geometric invariances. It would be of interest to evaluate its performance with rotation-invariant or scale-invariant curves to assess its ability to generate synthetic data that preserves these invariances while maintaining similarity to the original data and ensuring privacy.

Remaining within the realm of mobility applications, another future research direction aims to model trajectory data as a piecewise function and redefine the mathematical context for handling such functions. Specifically, with high-sampling rate GPS data, one would expect to describe movement in space adhering to the constraints of the road network. Instead of a smooth absolutely continuous function, the trajectory representation would be a sequence of piecewise continuous segments. For this approach, the modeling proposed by \citet{Steyer2023Elastic} can be referenced. 
High-frequency data are relatively rare because of accuracy issues and infrequent sampling. In these situations, before moving to the funcional modeling, an enrichment of the original data is needed. The first step would be to use map matching techniques to accurately position the data points on the road network. This process ensures that each GPS point is correctly aligned with the street map \citep{Yuan2010Interactive}. After map-matching, path enhancement techniques would be applied to refine the trajectory, improving its accuracy and continuity by interpolating between data points and smoothing out any irregularities \citep{Zheng2012Reducing, Li2021Trajectory}.

\section*{Acknowledgement and funding}
The authors thank Cuebiq Inc. for sharing the GPS dataset used in this work.
\\
The authors acknowledge the support from MUR, grant Dipartimento di Eccellenza 2023-2027.
\\
Arianna Burzacchi's work has been further supported by the Next Generation EU Programme REACT-EU through the PON Ph.D. scholarship “Development of innovative Eulerian privacy-preserving data analysis tools for designing more sustainable and climate-friendly human mobility services and infrastructures from high-resolution location data”.

\section*{Disclosure statement}
The authors report that there are no competing interests to declare.

\appendix
\section{Results of a different parameter choice} \label{sec:app1}

Instead of using the stabilization method, as discussed in Section \ref{sec:results}, the optimal values of $\alpha_0(K)$ are chosen here by setting a fixed threshold on the minimum distances. For a given threshold $b>0$, and for each $K$, $\hat{\alpha}_0(K)$ is identified as the first value where the indicator $\bm{I}^{(1)}$ exceeds this threshold $b$.
\begin{equation} \label{eq:}
    \hat{\alpha}_0(K) = \argmin_{\alpha>0} \{ \bm{I}^{(1)}[t_{\alpha}] > b \}\quad \forall K
\end{equation}
A threshold of $b=0.150$ is selected, corresponding to the $25$-th percentile of the distribution of distances between the original curves. This ensures that, for each synthetic-original pair of functions, at least $25$\% of the pairs between the original functions have a shorter distance. Since this threshold is lower than the stabilization value of the curves, the alternative method is expected to yield results that, while generally less compliant with privacy standards, might offer better utility preservation in some cases.

\begin{figure}[tbh]
    \centering
    \includegraphics[width=8cm]{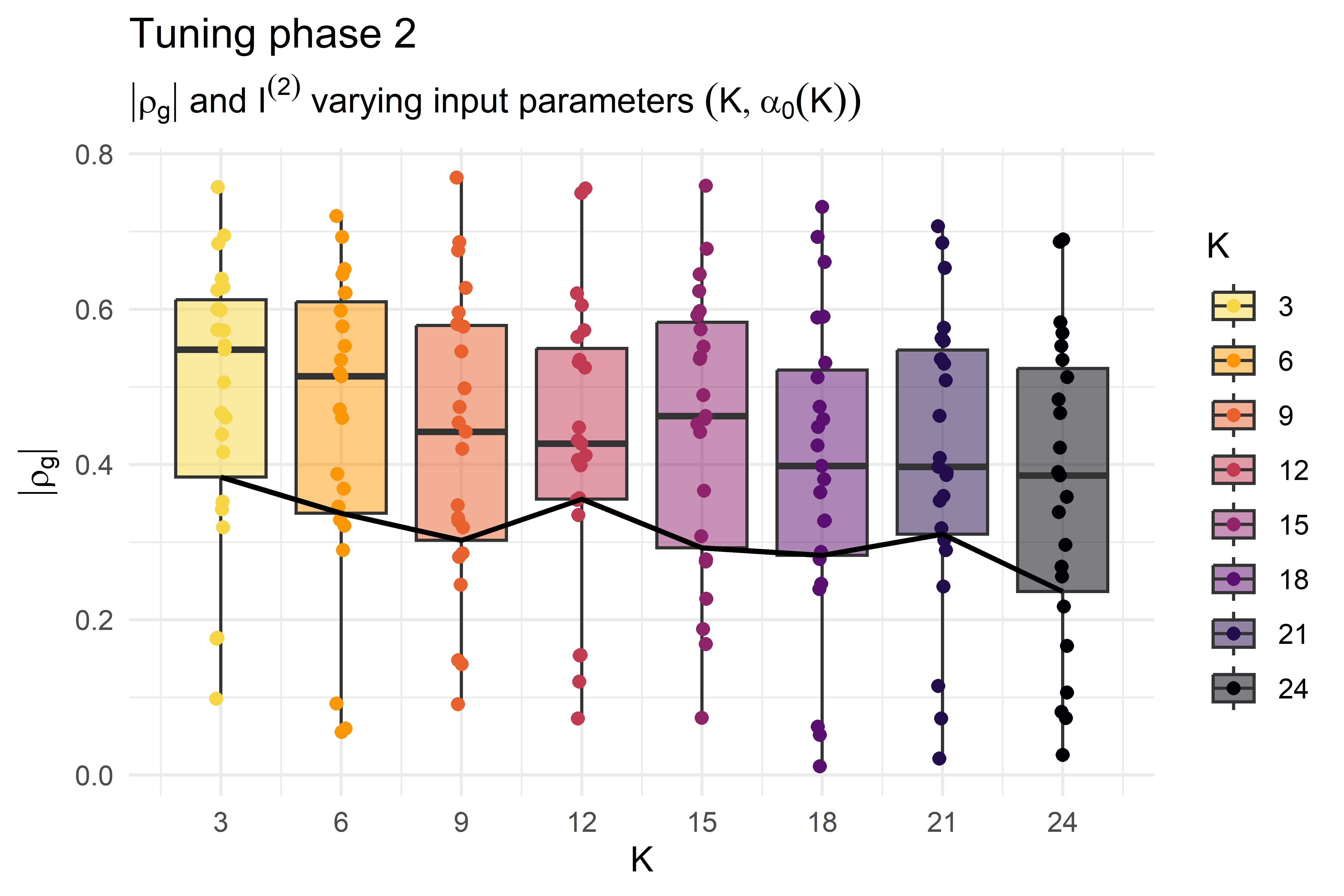}
    \caption{Boxplot of the absolute correlations of distances within groups, output of the second phase of tuning procedures for the choice of $\hat{K}$, given $\hat{\alpha}_0(K)$ as in Table \ref{tab:tuning:2}. The straight black line depicts the variation of the indicator $\bm{I}^{(2)}$ with $K$.}
    \label{fig:tuning_alpha_k:2}
\end{figure}

Following this criterion, the optimal values $\hat{\alpha}_0(K)$ for each $K$ are derived and listed in Table \ref{tab:tuning:2}. With these combinations of $K$ and $\hat{\alpha}_0(K)$, the distance correlations and the indicator vector $\bm{I}^{(2)}$ are computed. This time, the maximum is reached with $K=3$ and the corresponding $\alpha_0=7$, as illustrated in Figure \ref{fig:tuning_alpha_k:2}. The final optimal parameters used in this case will therefore be $K=3$ and $\alpha_0=7$. Compared to the previous case, the same $\alpha_0$ value is maintained, but with a smaller $K$, which is expected to generate results more closely aligned with specific observations and less influenced by smoothing effects, as would be the case with the larger $K=6$.

\begin{table}[tbh]
    \centering
    \begin{tabular}{|l|cccccccc|}
    \hline
        $K$ & 3 & 6 & 9 & 12 & 15 & 18 & 21 & 24 \\
        \hline
        $\hat{\alpha}_0(K)$ & 7 & 3 & 3 & 3 & 3 & 3 & 3 & 3 \\
    \hline
    \end{tabular}
    \caption{Optimal values of $\alpha_0$ for each value of $K$. The results are reported when using the threshold method, i.e., by chosing $\hat{\alpha}_0(K)$ as the first point after the privacy threshold is exceeded.}
    \label{tab:tuning:2}
\end{table}

\begin{figure}[tbh]
    \centering
    \subfigure[Original (sample A)]{\includegraphics[width=4cm]{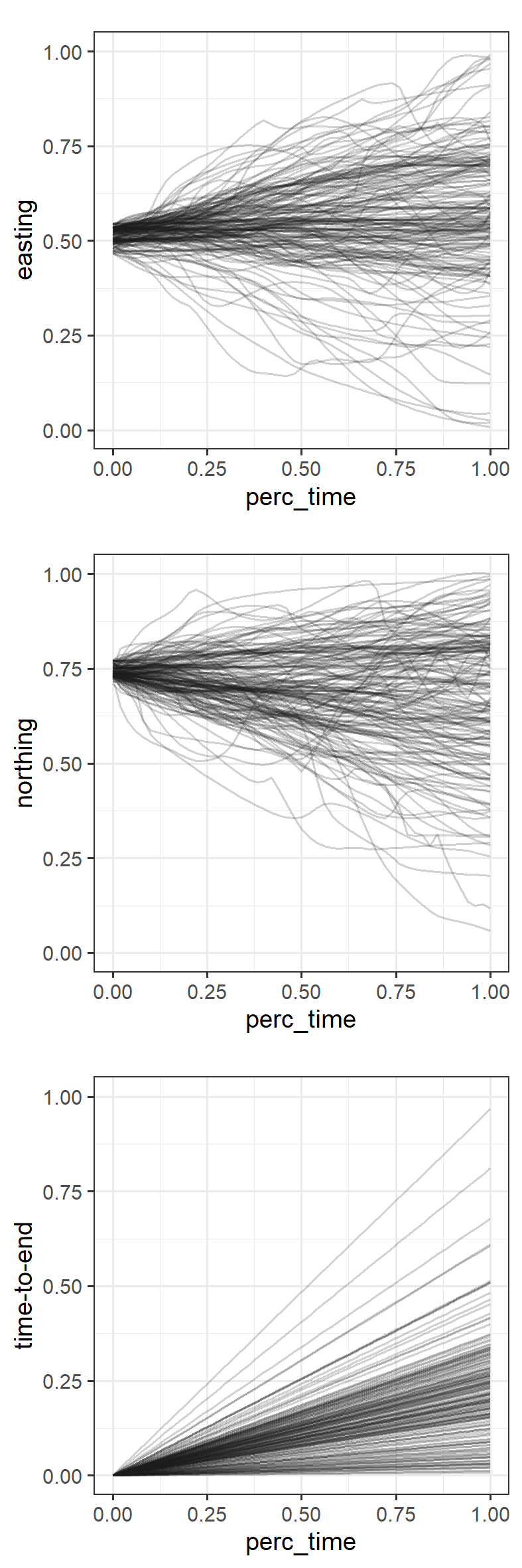}}
    \hspace{0.25cm}
    \subfigure[Synthetic (sample A)]{\includegraphics[width=4cm]{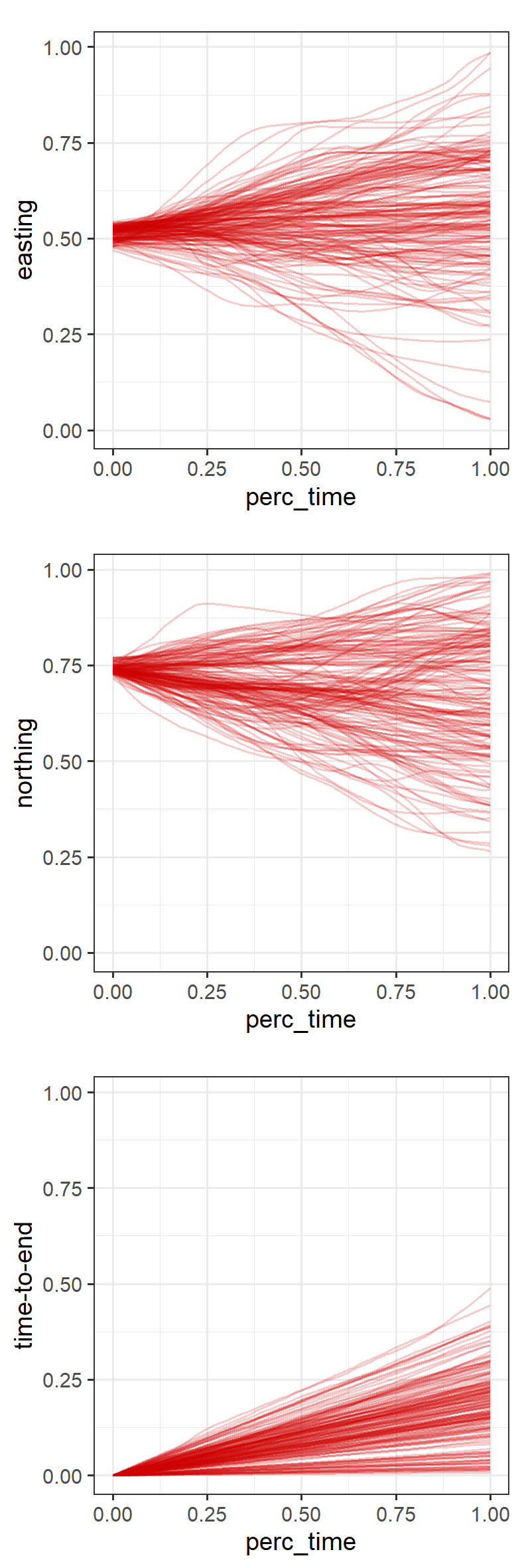}}
    \hspace{0.25cm}
    \subfigure[Synthetic (sample B)]{\includegraphics[width=4cm]{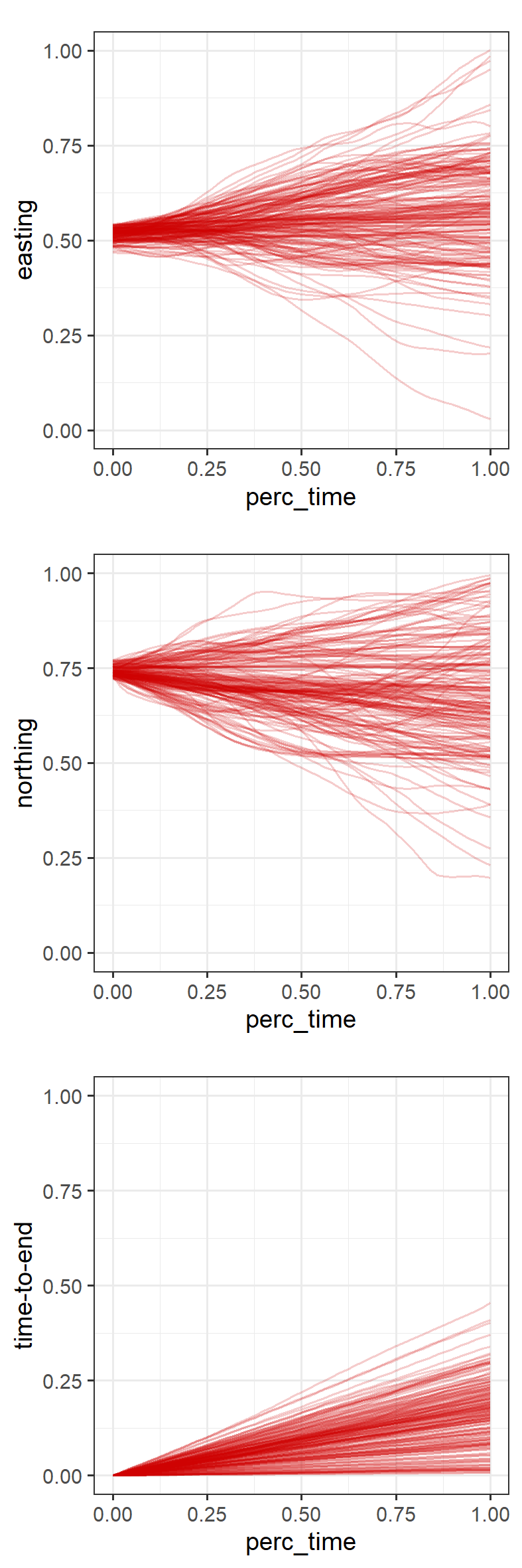}}
    \caption{Representation of some functions, both original (left) and synthetic (center and right), produced when using $K=3$ and $\alpha_0=7$ in the FDASynthetsis method.}
    \label{fig:viz:functional:2}
\end{figure}

\begin{figure}[tbh]
    \centering
    \subfigure[Original (sample A)]{\includegraphics[width=4cm]{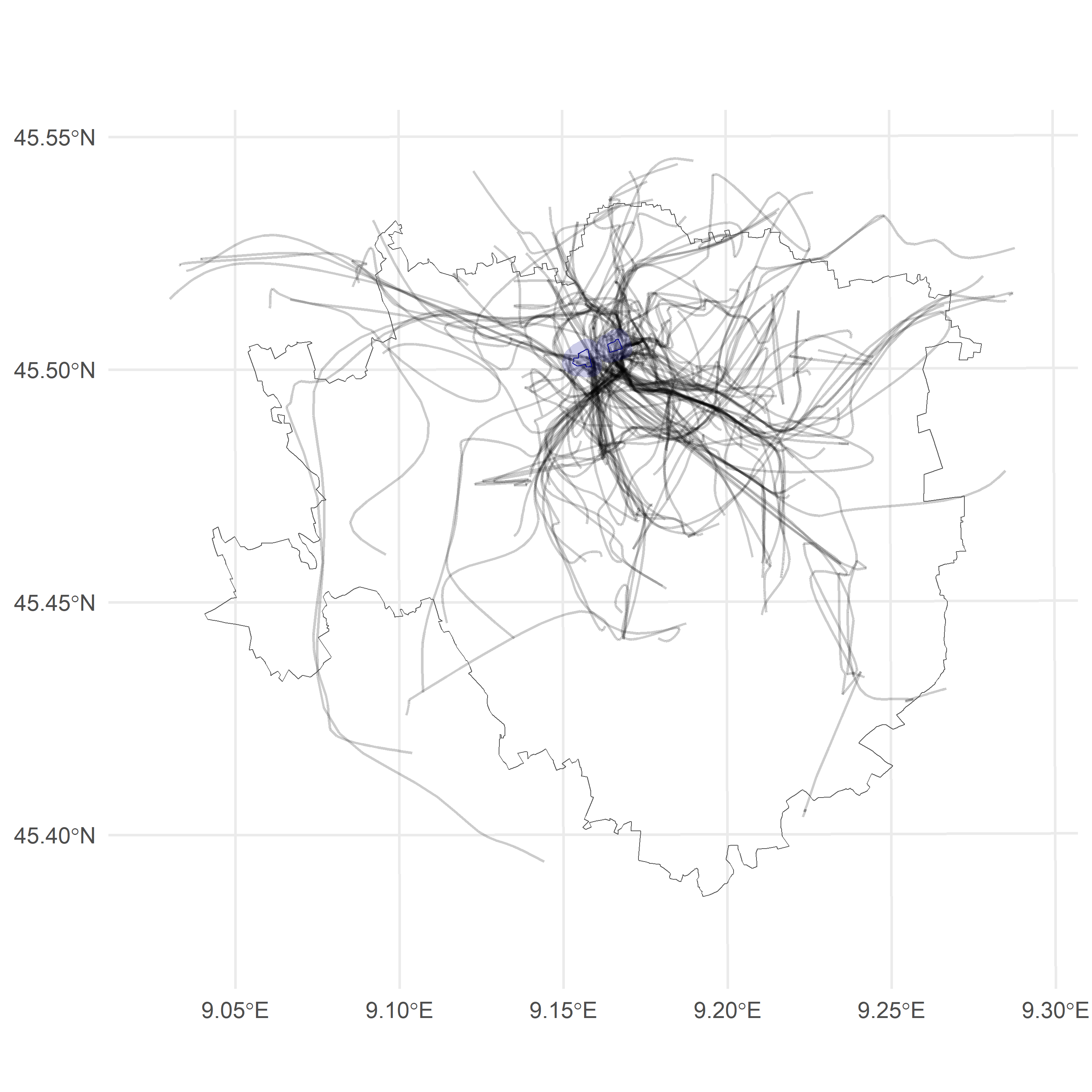}}
    \hspace{0.5cm}
    \subfigure[Synthetic (sample A)]{\includegraphics[width=4cm]{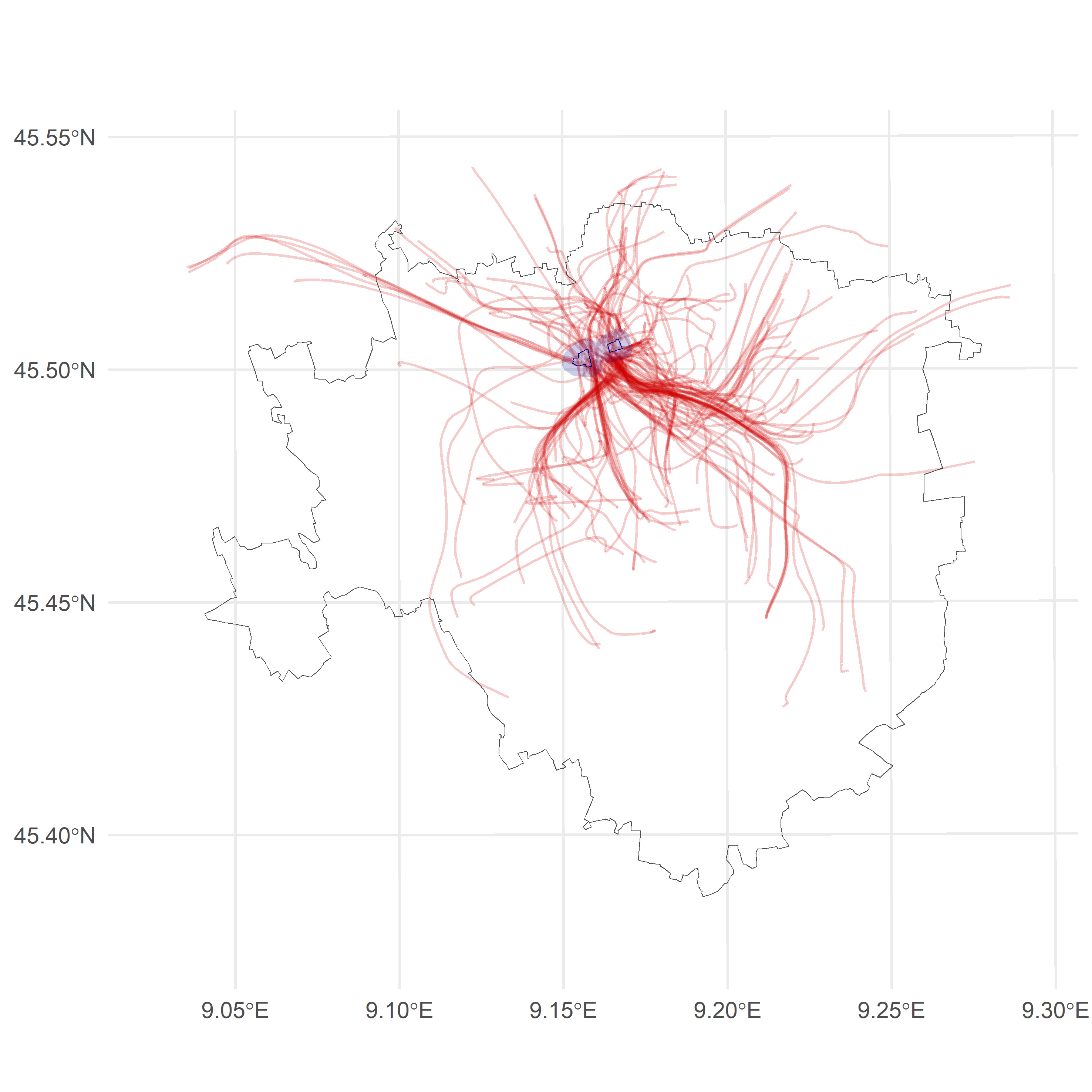}}
    \hspace{0.5cm}
    \subfigure[Synthetic (sample B)]{\includegraphics[width=4cm]{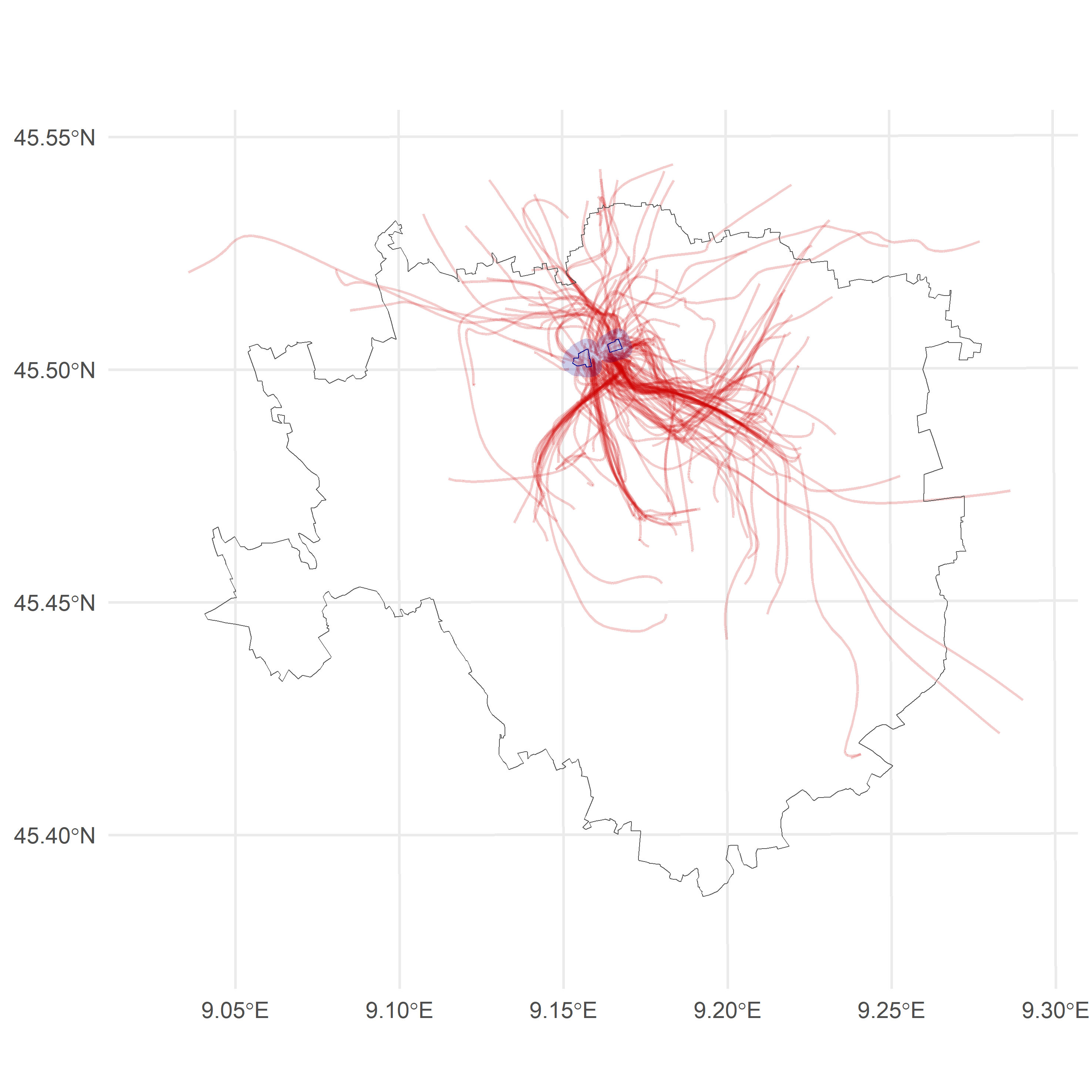}}
    \caption{Projection on the original map of Milan of some functions, both original (left) and synthetic (center and right), produced when using $K=3$ and $\alpha_0=7$ in the FDASynthetsis method. The Bovisa campus is highlighted in blue. The black border represents the boundary of the Milan area.}
    \label{fig:viz:map:2}
\end{figure}

\begin{figure}[tbh]
    \centering
    \subfigure[Original (sample A)]{\includegraphics[width=4cm]{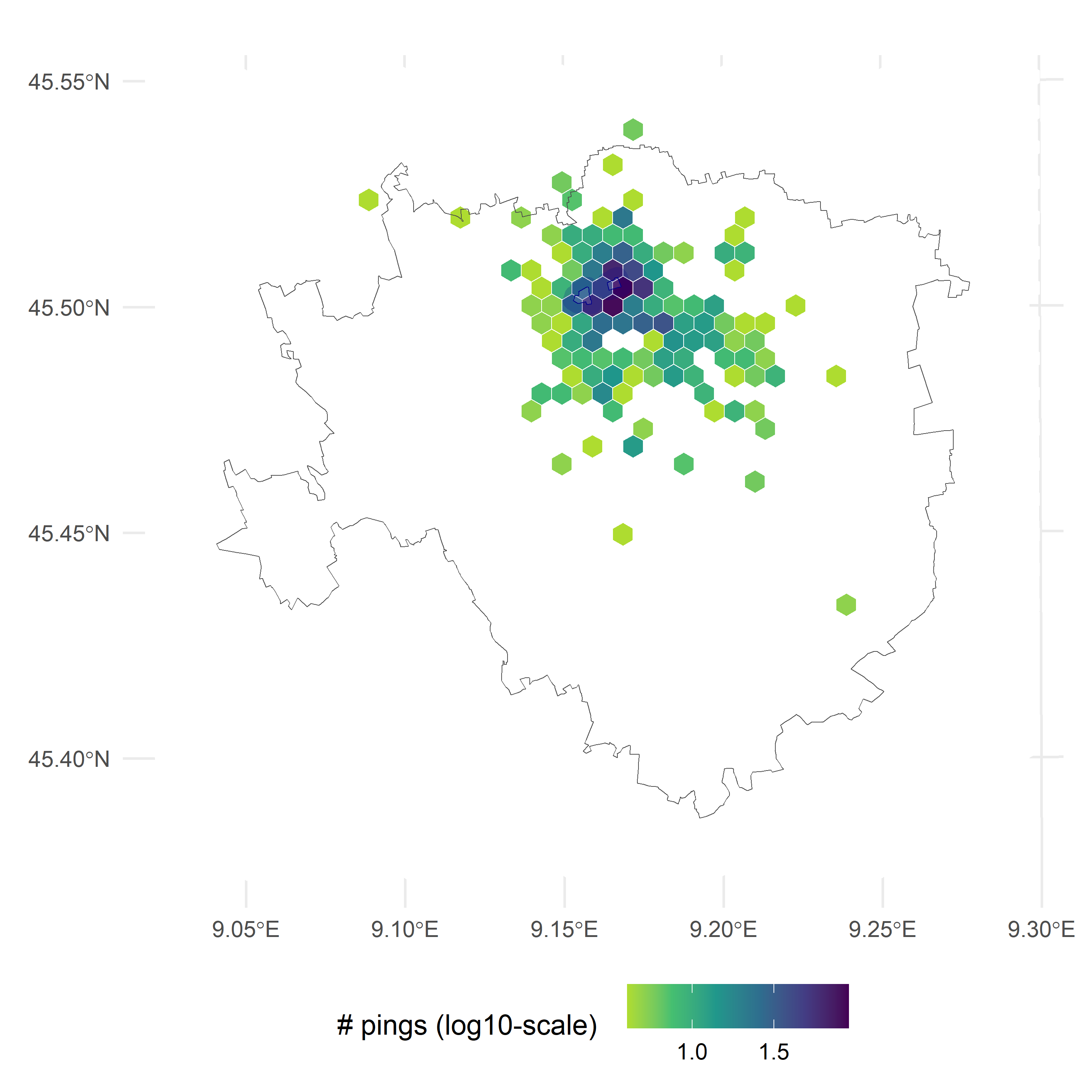}}
    \hspace{0.5cm}
    \subfigure[Synthetic (sample A)]{\includegraphics[width=4cm]{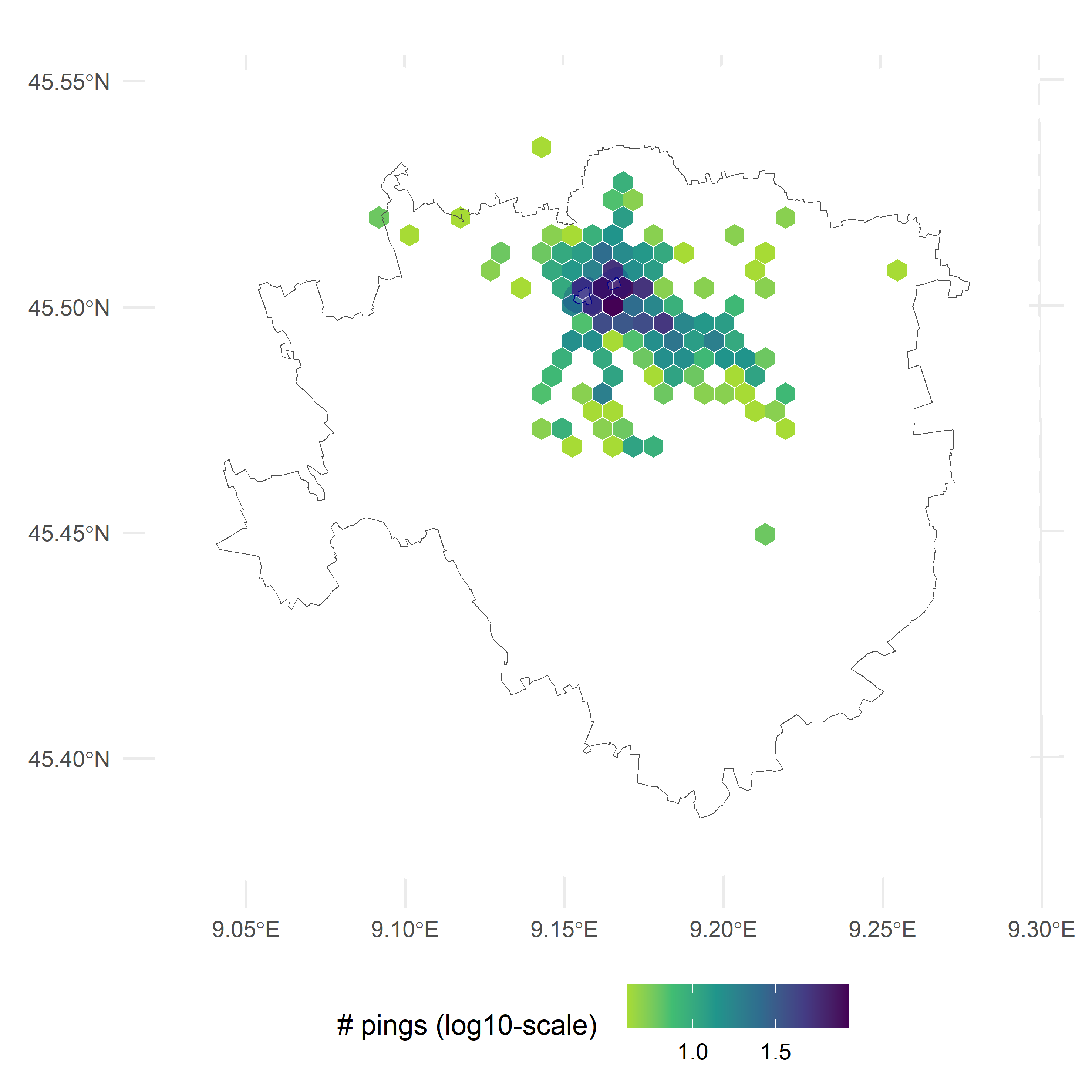}}
    \hspace{0.5cm}
    \subfigure[Synthetic (sample B)]{\includegraphics[width=4cm]{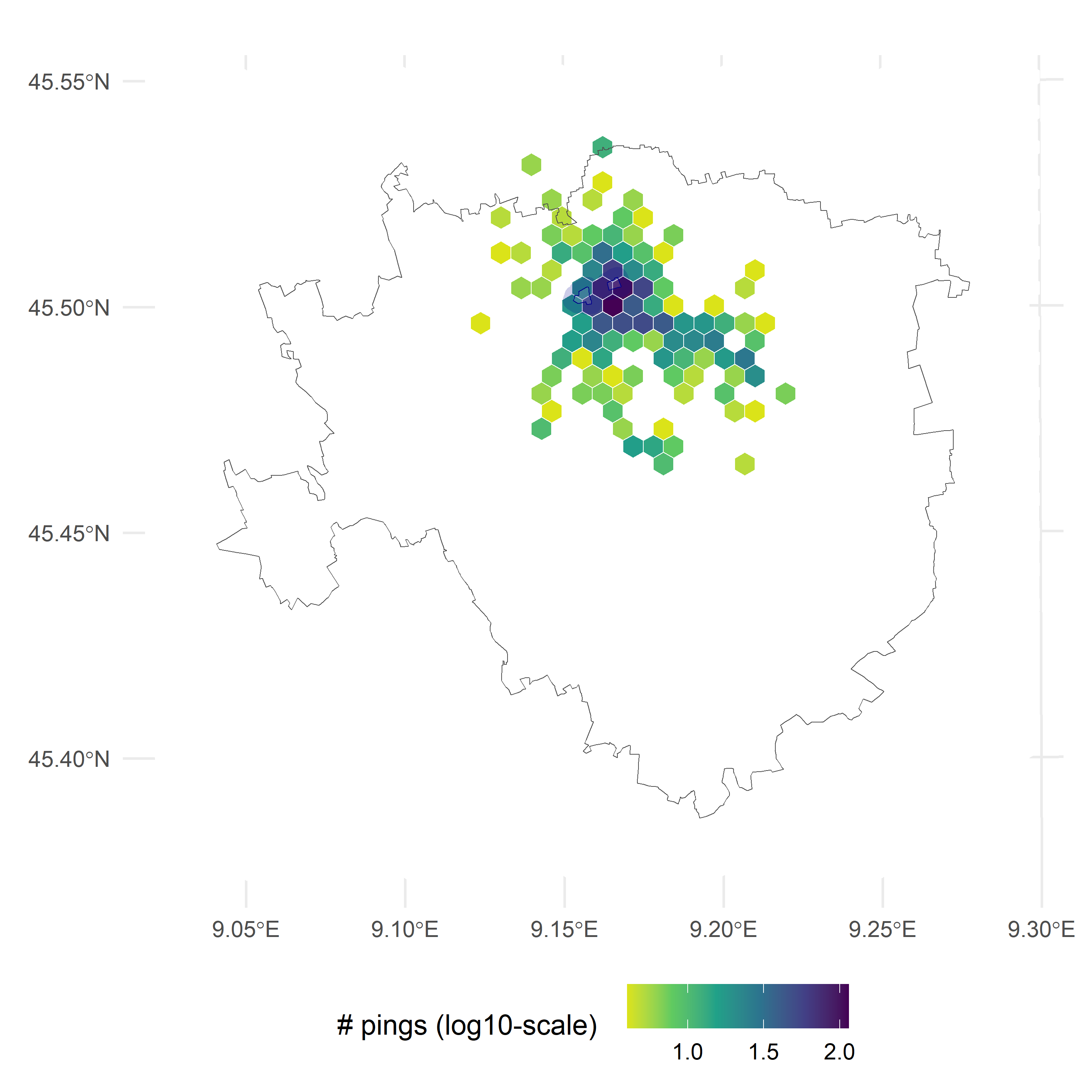}}
    \caption{Heatmaps of the most visited neighbors after subsampling. Synthetic curves are produced using $K=3$ and $\alpha_0=7$ in the FDASynthetsis method.}
    \label{fig:viz:heatmap:2}
\end{figure}

The same visual inspections and quantitative analyses are conducted to evaluate the synthetic data generated using the new optimal combination of parameters. Figures \ref{fig:viz:functional:2}, \ref{fig:viz:map:2}, and \ref{fig:viz:heatmap:2} show samples of original curves and synthetic curves, respectively, as multidimensional functions, then reprojected on the physical map, then discretized via subsampling to check the visit distributions. 
From these figures, it is evident that the synthetic data preserves the structure of the original functional data. The synthetic curves follow trends similar to the real trajectories, maintaining the overall charachteristics of the spatiotemporal dynamics (Figure \ref{fig:viz:functional:2}). They retain realistic mobility patterns that are consistent with those observed in the original data, closely following the main paths of the original trajectories (Figure \ref{fig:viz:map:2}). Finally, the method is able to generate data that reflects mobility dynamics on an aggregate scale, as demonstrated by the comparable spatial distribution in terms of visit frequency (Figure \ref{fig:viz:heatmap:2}). Compared to the previously analyzed case, the current approach maintains greater variability in the data. Indeed, these visualizations show that even rarer behaviors are captured in the synthetic dataset, such as the presence of longer trajectories from the south of the city, and that visits are distributed over a larger area covering more cells.

\begin{figure}[tbh]
    \centering
    \subfigure[Test for the mean functions]{\includegraphics[width=6.5cm]{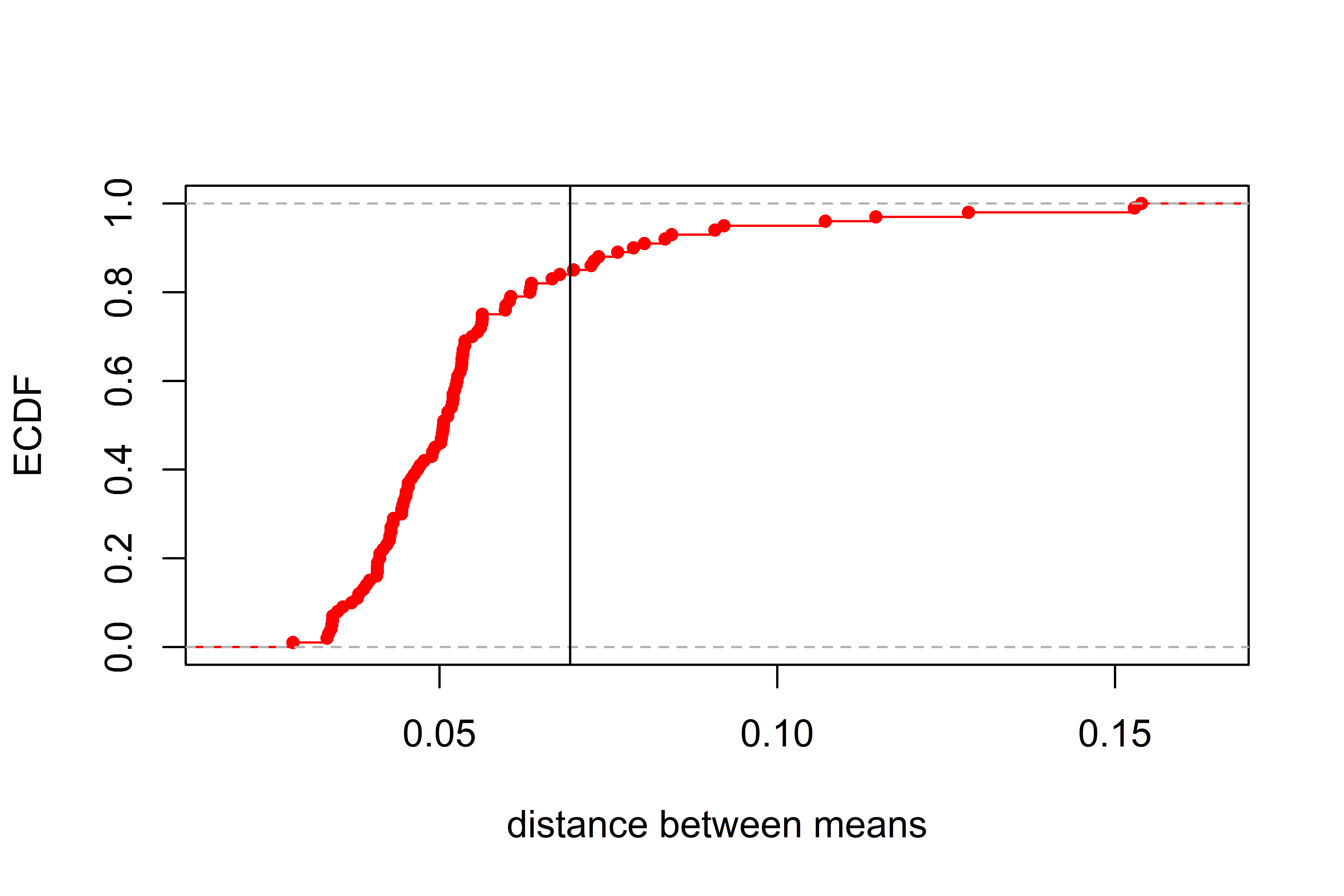}}
    \hspace{0.5cm}
    \subfigure[Test for the covariance operators]{\includegraphics[width=6.5cm]{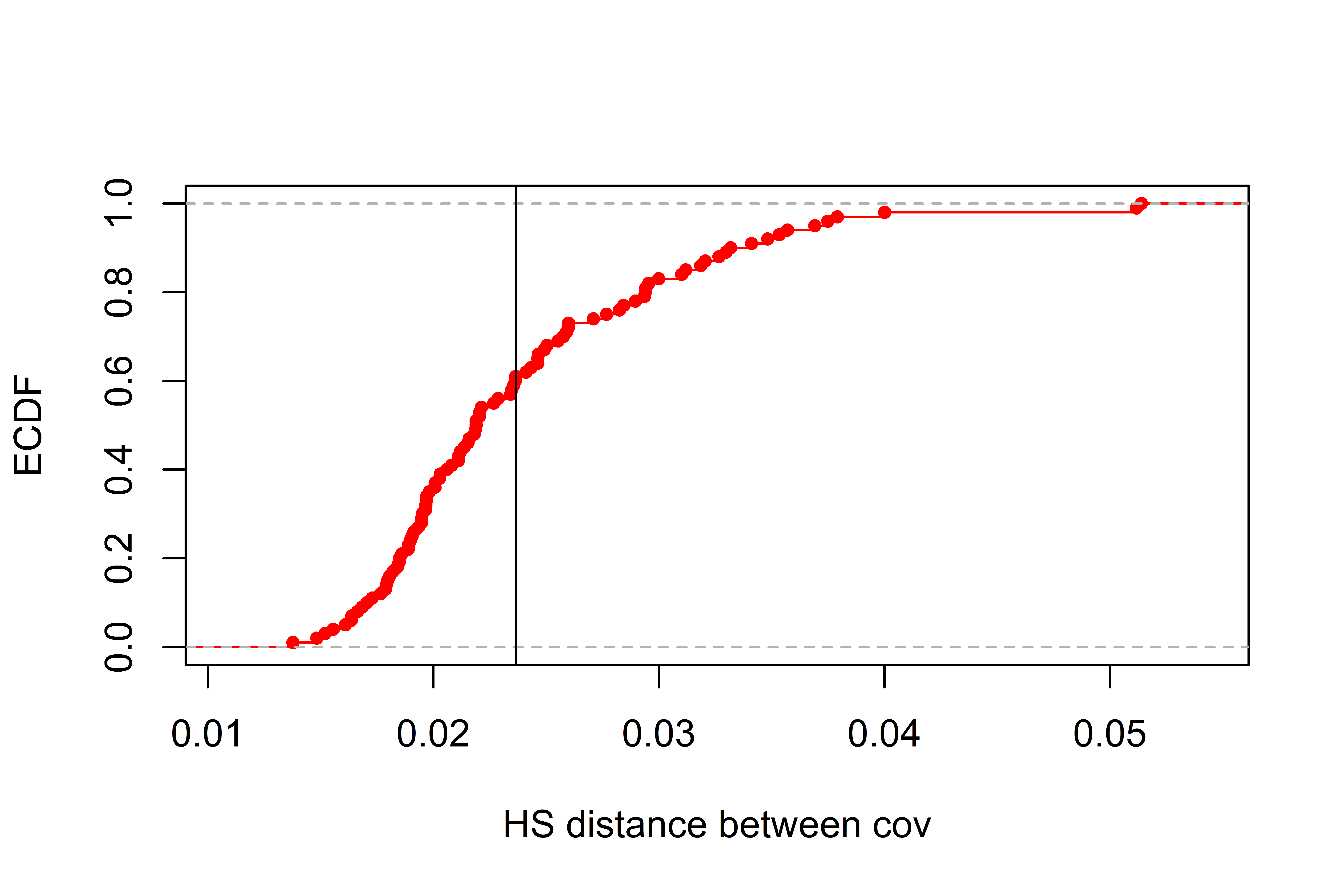}}
    \caption{ECDFs of the test statistics evaluating the method when $K=3$ and $\alpha_0=7$. On the left, ECDF of the distances between mean functions. On the right, ECDF of the distances between covariance operators. In both plots the vertical line represents the observed test statistic.}
    \label{fig:test2}
\end{figure}

In addition to the visualizations, permutation tests described by the hypotheses in Equations \ref{eq:test1} and \ref{eq:test2} are applied. Following the same methodological approach as before, the equality of the mean functions of the two datasets is tested. By using $500$ random permutations of the labels within the dataset, the ECDF is obtained, as shown in Figure \ref{fig:test2}(a), and results in a p-value of $p_{\mu} = 0.16$. 
Subsequently, the equality of the covariance operators is tested, using the Hilbert-Schmidt distance between operators as the test statistic. After $500$ iterations of the permutation test, the resulting ECDF, depicted in Figure \ref{fig:test2}(b), leads to a p-value of $p_{\Sigma} = 0.39$. In both tests, the p-values are not low enough to reject the null hypotheses, thereby supporting the equality of the mean functions and the equality of the covariance operators.

\begin{figure}[tbh]
    \centering
    \includegraphics[width=6cm]{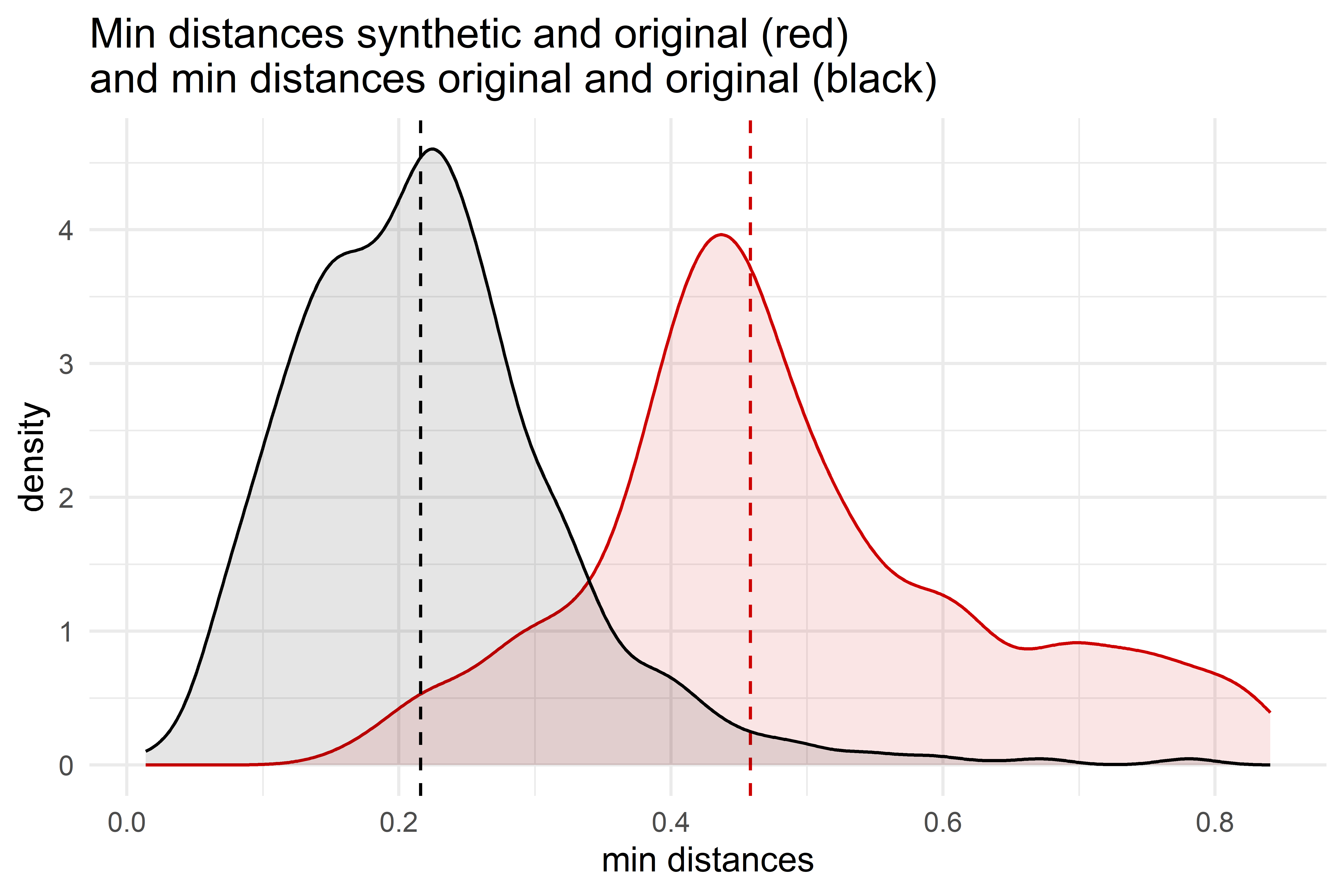}
    \caption{Distribution of the minimum distances among original functions (in black) and between original and synthetic curves (in red) when $K=3$ and $\alpha_0=7$. The vertical dotted lines represent the value of the median of the distributions.}
    \label{fig:privacy_check:2}
\end{figure}

Using a smaller minimum distance to control privacy, one needs to check the absence of 1-to-1 correspondences between synthetic and original data. Figure \ref{fig:privacy_check:2} shows the distribution of the minimum distances between synthetic and original functions (red) and among originals (black). In this case as well, the mean distances are large enough to support the preservation of privacy.

\newpage
\bibliographystyle{abbrvnat}
\bibliography{biblio}

\end{document}